\documentclass[authoryear,review,5p,times]{elsarticle}
\usepackage{url} 
\usepackage{doi} 
\usepackage{xcolor}
\usepackage{CJKutf8}
\usepackage{threeparttable}

\usepackage{amssymb}
\usepackage{amsmath}
\usepackage{booktabs}
\usepackage{lineno}
\journal{Journal of High Energy Astrophysics}

\begin{document}

\begin{frontmatter}



\title{Statistical properties of type-II bursts in Rapid Burster observed with {\it Insight}-HXMT}

\author[1]{Shi-Peng Wang}
\author[1,2]{Long Ji}
\affiliation[1]{organization={School of Physics and Astronomy, Sun Yat-Sen University},
            addressline={DaXue Road 2},
            city={Zhuhai},
            postcode={519082},
            state={Guangdong},
            country={People's Republic of China}}
\affiliation[2]{organization={CSST Science Center for the Guangdong-Hong Kong-Macau Greater Bay Area},
            addressline={DaXue Road 2},
            city={Zhuhai},
            postcode={519082},
            state={Guangdong},
            country={People's Republic of China}}
\ead{jilong@mail.sysu.edu.cn}




\begin{abstract}
We report the detection of type-II X-ray bursts in a neutron star low mass X-ray binary MXB 1730-335 (also known as Rapid Burster) observed with Insight-{\it HXMT} satellite.
We found significant variations in the burst properties along with the decay phases of its 2017 and 2020 outbursts, such as switches between different burst modes.
We investigated the statistical distributions of burst parameters (peak flux $F_{\rm peak}$, fluence $E$, burst duration $t_{\rm dur}$, recurrence time $\Delta t$) and studied their correlations.
We confirm the relaxation oscillator behavior ($E \propto \Delta t^{\alpha}$), and find that for mode-2 bursts the index $\alpha$ is around 1. For mode-1 bursts, $\alpha$ varies with $\Delta t$ and the $E-\Delta t^{\alpha}$ relation can be described by a broken powerlaw model.
We also confirm that $F_{\rm peak}$ is correlated with $E$ below a critical value and is independent of $\Delta t$.
\end{abstract}




\end{frontmatter}

\begin{CJK*}{UTF8}{gbsn}
\end{CJK*}

\section{Introduction}
Low mass X-ray binaries (LMXBs) are celestial systems composed of a compact object and a donor star with a mass $\lesssim$ 1 solar mass ($M_{\odot}$). 
The compact star can be either a black hole or a neutron star, dividing LMXBs into two categories.
In such systems, mass is transferred from the Roche-lobe-filling companion star to an accretion disc, before subsequently accreting onto the compact object.
Numerous X-ray bursts have been observed in neutron star LMXBs, characterized by a sudden and significant increase in X-ray intensity. 
These bursts can be further categorized into two types based on their energy sources \citep[for reviews see, e.g., ][]{1993Lewin,2004Cumming,2006Strohmayer,2021Galloway}:
1) type-I X-ray bursts occur when matter accretes and accumulates on the neutron star surface, triggering unstable thermonuclear explosions;
2) type-II X-ray bursts probably originate from the release of gravitational energy when the accreting matter suddenly falls onto the neutron star \citep{1978Hoffman}, although there are still no widely accepted models that can fully explain the behavior of type-II X-ray bursts.
{To date, type-I X-ray bursts have been detected in more than 100 sources, whereas only two sources, the Bursting Pulsar \citep[BP, also known as GRO J1744-28,][]{1996Kouveliotou.C.} and Rapid Burster \citep[RB, also known as MXB 1730-335,][]{1976Lewin}, exhibit type-II X-ray bursts.}

In this study, we focus on the Rapid Burster (hereafter RB), a source situated in the globular cluster Liller 1 at a distance of $\sim$ 7.9\,kpc \citep{2010Valenti}.
Its name comes from the repetitive bursting behavior \citep{2002Masetti}.
Compared with the Bursting Pulsar (BP), which presents only type-II bursts, the RB displays both type-I and -II bursts.
In some cases, type-I bursts are predominant and no type-II bursts are observed when the source is bright.
Conversely, there are outburst states characterized by the dominance of type-II bursts with only a few type-I bursts.
Type-II bursts show a variety of morphological modes that may suddenly switch between each other: 1) mode-0 is characterized by long, intense, and flattened bursts; 2) mode-1 features a series of short, rapidly recurring bursts, followed by a relatively long and energetic burst at the end; and 3) mode-2 consists entirely of short, rapidly recurring bursts \citep{1999.R.Guerriero,1979Marshall,1993Lewin}.
After long bursts in mode-0 and mode-1, the persistent emission rises to a higher level before the next burst, forming the so-called ``hump" feature, which has never been observed following short bursts \citep{2015.T.Bagnoli}.
The most significant property of type-II bursts in RB is the $E-\Delta t$ relation (also known as the relaxation oscillator relation), where $E$ is the burst fluence and $\Delta t$ is the recurrence time.
This relationship implies a mass storage mechanism, suggesting that after a stronger burst, a longer recurrence period will be needed to trigger the subsequent burst, due to the emptied reservoir that needs to be refilled with mass.
The specific triggering mechanism is still unknown although several models have been proposed based on magnetospheric and thermal instabilities \citep[e.g.,][]{1993Lewin,1993Spruit,2015.T.Bagnoli}.
The broadband spectra of type-II bursts have been studied using {\it Insight}-HXMT observations performed in August 2017 \citep{Chen2021}.
In this study, using the same dataset and additional observations in 2020, we aim to investigate their statistical properties.
This paper is structured as follows: In Sections~\ref{Observations}, ~\ref{Burst_identification} and~\ref{Results} we describe on the data analysis and present the results.
We compare with previous observations 
in Section~\ref{Discussion}.

\begin{figure*}[htbp]
\centering
\includegraphics[width=1\textwidth]{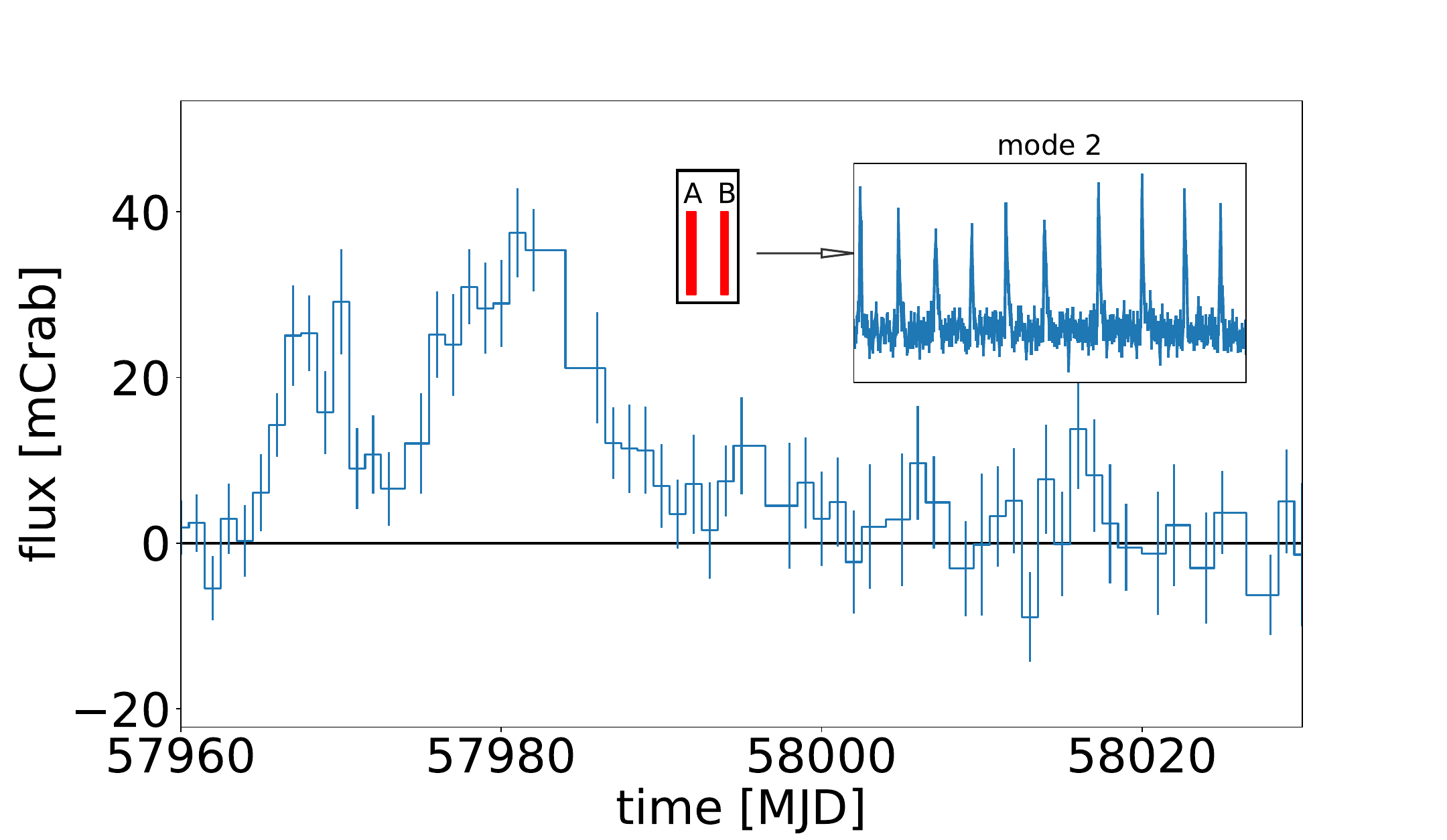}
\includegraphics[width=1\textwidth]{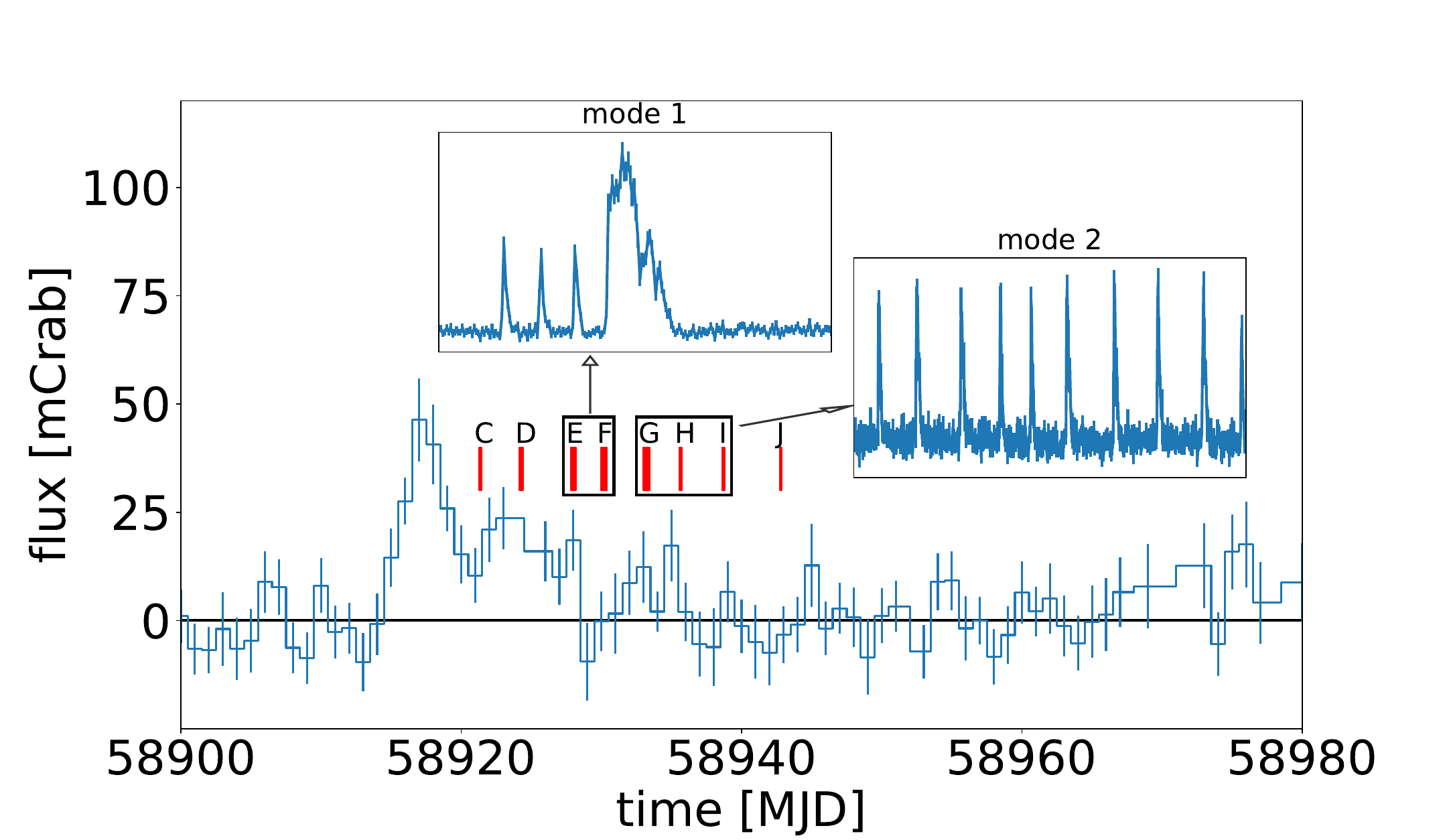}
\centering
\caption{Daily lightcurves of the RB in 2017 (top) and 2020 (bottom) observed with {\it Swift}/BAT in the energy band of 15-50\,keV.
The vertical red lines mark the epochs when {\it Insight}-HXMT observations were conducted.
The horizontal black line marks the zero value.
The insets show representative burst profiles: mode-1 bursts observed in epochs E–F, and mode-2 bursts observed in epochs A–B, G–I.
No bursts were detected in Epochs C, D and J.
}

\label{fig:BAT}
\end{figure*}

\begin{figure}[htbp]
\centering
\includegraphics[width=0.5\textwidth]{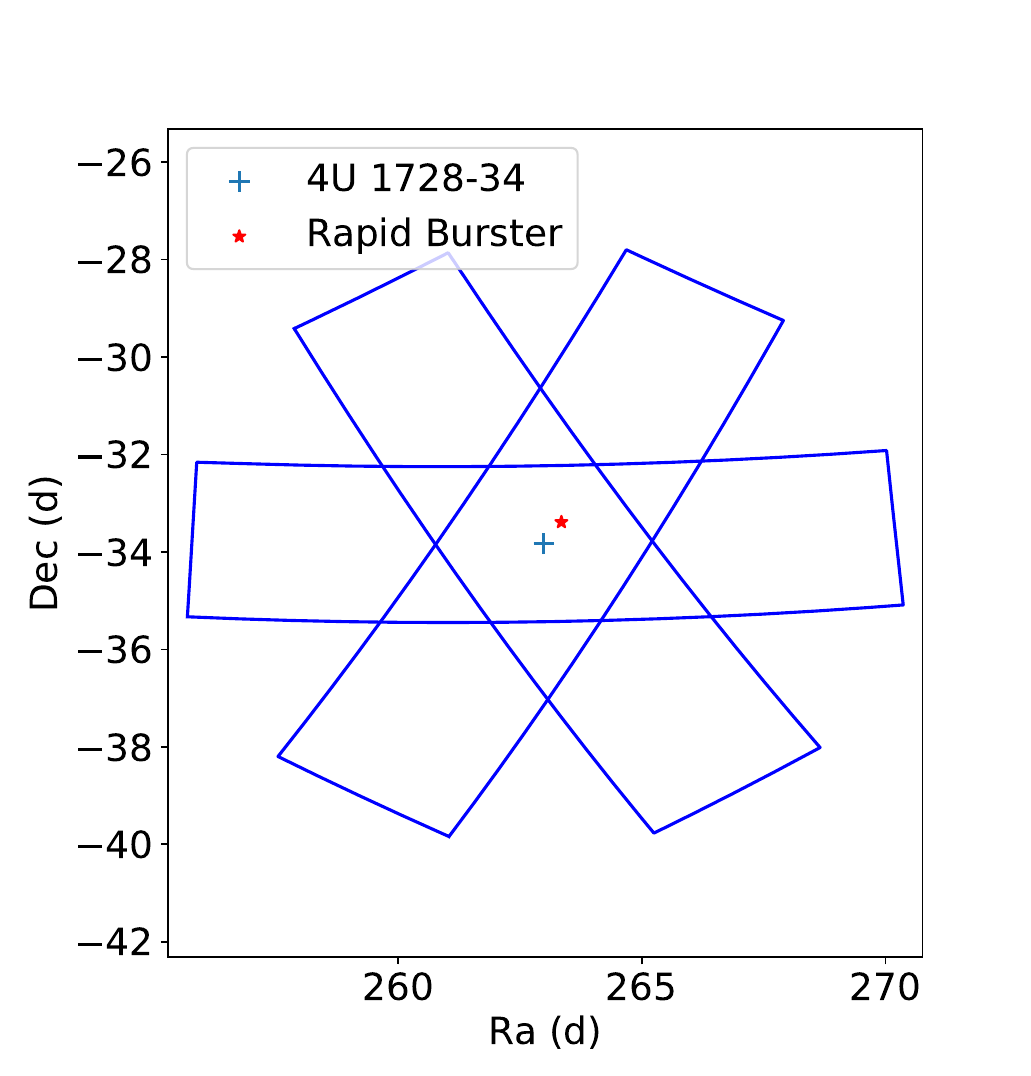}

\caption{A representative configuration of LE collimators' FoVs.
Our target (red star) and the contamination source (blue ``+") are close to each other.  
}
\label{fig:FOV}
\end{figure}

\section{Observation and data reduction}
\label{Observations}
\label{HXMT}
\begin{table*}
        \centering
        \caption{
        A summary of {\it Insight}-HXMT observations, where $n_{\rm burst}$ is the number of type-II bursts detected. 
}

    \begin{tabular}{ccccccc}
        \toprule
        Epochs  & ObsID & Start time (UTC) & Stop time (UTC) & Exposure (s) & n\textsubscript{burst} & Bursting rate (${hour}^{-1}$)\\
        \hline
        A & P0101303001 & 2017-08-26T13:45:18 & 2017-08-27T04:24:03 & 11272 & 76 & 24.3 \\
        B & P0101303002 & 2017-08-28T16:43:26 & 2017-08-29T03:55:45 & 8502 & 223 & 94.4 \\
        \hline
        C & P0214064001 & 2020-03-13T02:33:12 & 2020-03-13T12:10:50 & 6073 & 0 & 0 \\
        D & P0214064002 & 2020-03-16T00:31:00 & 2020-03-16T10:20:55 & 6995 & 0 & 0 \\
        E & P0214064003 & 2020-03-19T19:09:18 & 2020-03-20T04:47:54 & 12948 & 108 & 30.0 \\
        F & P0214064004 & 2020-03-21T23:38:13 & 2020-03-22T09:22:59 & 9280 & 100 & 38.8 \\
        G & P0214064005 & 2020-03-24T23:13:03 & 2020-03-25T10:39:51 & 10674 & 170 & 57.3 \\
        H & P0214064006 & 2020-03-27T13:16:21 & 2020-03-27T18:09:01 & 4488 & 62 & 49.7 \\
        I & P0214064007 & 2020-03-30T14:29:18 & 2020-03-30T19:21:07 & 3651 & 21 & 20.7 \\
        J & P0214064008 & 2020-04-03T17:09:54 & 2020-04-03T22:00:56 & 4953 & 0 & 0 \\
    \hline
    \end{tabular}
    \label{tab:observations}
\end{table*}

In this study, we studied two {\it Insight}-HXMT datasets obtained in 2017 and 2020.
The detailed information is summarized in Table~\ref{tab:observations}.
RB is a transient source exhibiting outbursts with a duration of several tens of days \citep{1999.R.Guerriero}.
Figure~\ref{fig:BAT} presents 15-50\,keV long-term lightcurves monitored by {\it Swift}/BAT, which indicates that our observations in both 2017 and 2020 were conducted during the decay phase of outbursts at different flux levels.
This allows us to study burst behaviors along with outburst evolutions at different accretion rates.
In the following, we define these observations as Epochs A-J, and type-II bursts were only found in A-B and E-I.

{\it Insight}-HXMT is equipped with three slat-collimated instruments operating in the energy range of 1–250\,keV \citep{Zhang.S..2014,Zhang.2020}, including the high energy telescope (HE; 20–250\,keV), the medium energy telescope (ME; 5–30\,keV) and the low energy telescope (LE; 1–10\,keV) \citep{Liu.Cong2020, Cao.Xue2020, Chen.Yong2020}.
Since bursts are more significant in the low energy band, we used LE only in this paper.

The data reduction is conducted using the official software (HXMTDAS\footnote{\url{http://hxmtweb.ihep.ac.cn/software.jhtml}}) v2.06 with the calibration database CALDB 2.07. 
The good time intervals (GTI) were selected according to the following criteria: (1) an elevation angle exceeding $8^{\circ}$; (2) a geomagnetic cutoff rigidity threshold $>$\,6 GV; and (3) exclusion of South Atlantic Anomaly (SAA) passages to avoid instrumental noise contamination. 
%
Compared with the previous study by \citet{Chen2021} who used the default GTI selection (an elevation angle $>$ $10^{\circ}$; a geomagnetic cutoff rigidity threshold $>$\,8\,GV and exclusion of additional 300\,s of data before and after the SAA passages), our filtering criteria are much looser. 
This is due to the fact that our aim is to obtain a more complete burst sample while the accurate estimation of the instrumental background is not necessary.
A side effect of non-standard GTI filtering is that at passages close to SAA or the bright Earth limb, some artificial flares can be produced by the high background at the start or end of consecutive GTI segments, which appear similar to truncated bursts.
Therefore, in this study, we excluded all these suspicious flares and only studied bursts with complete shapes.
The total exposure was approximately 20\,ks and 60\,ks for the observations in 2017 and 2020, respectively.

\begin{figure}[htbp]
\centering
\includegraphics[width=0.5\textwidth]{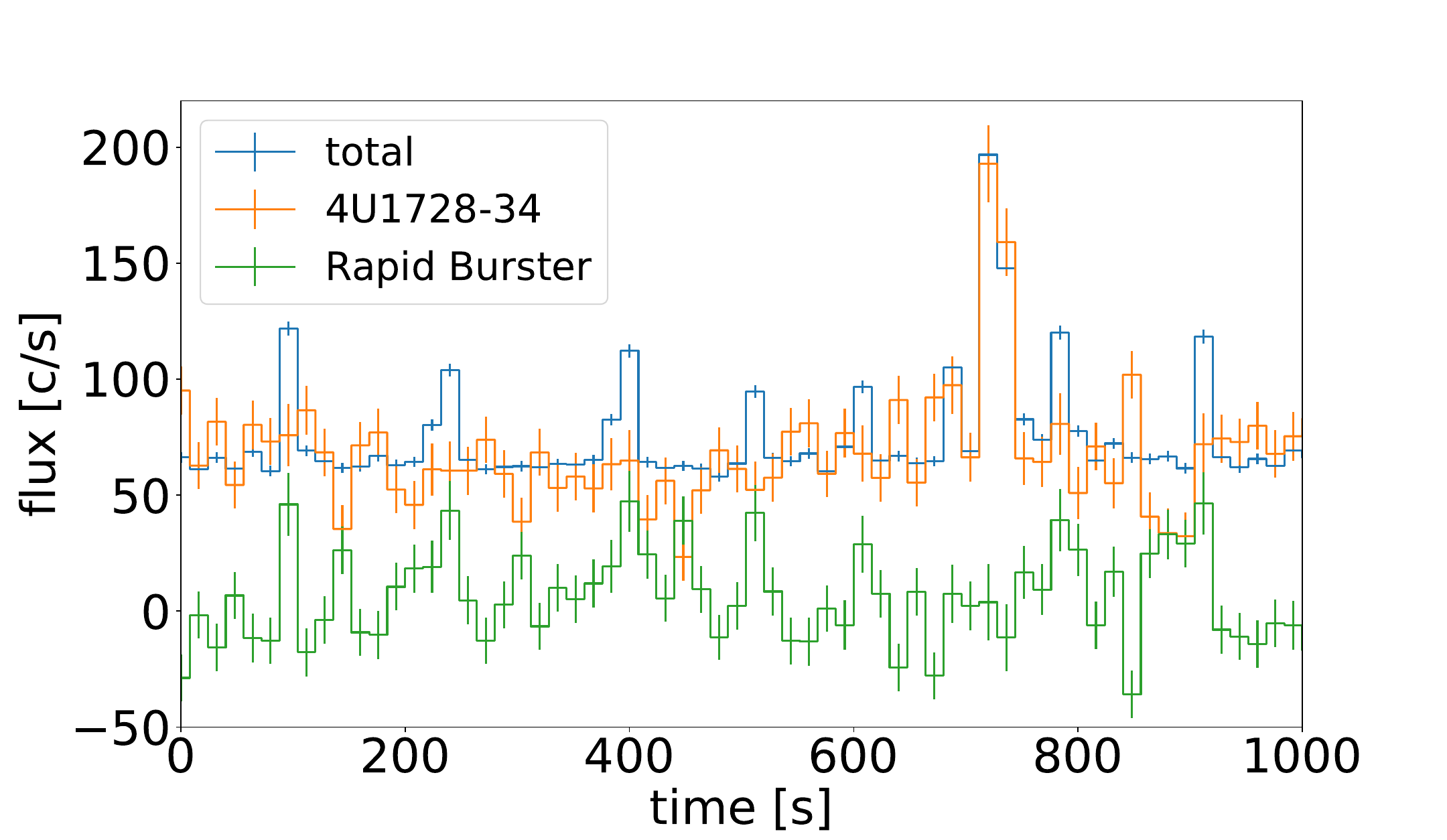}
\includegraphics[width=0.5\textwidth]{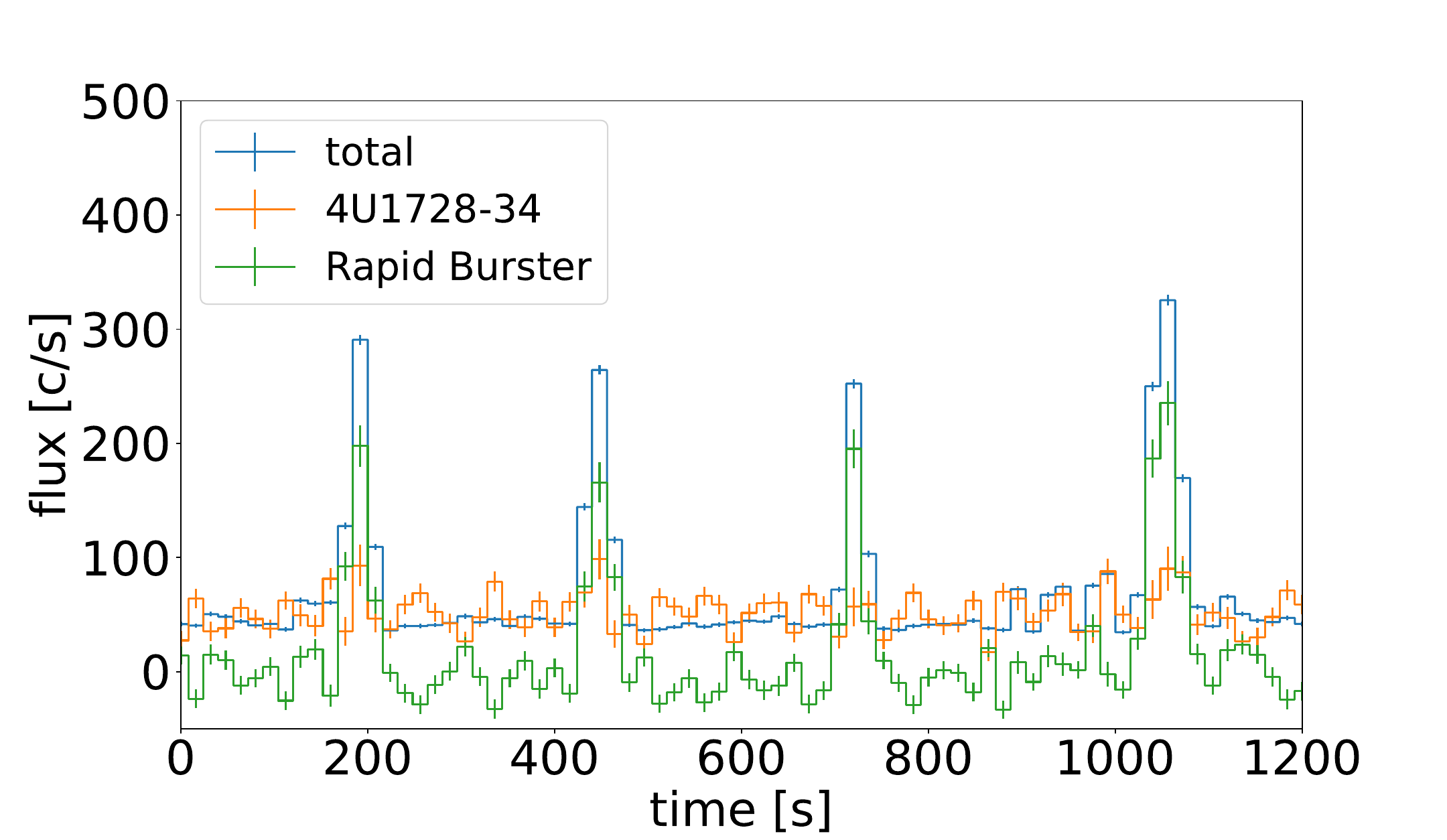}
\centering
\caption{Decompositions of background-subtracted lightcurves (blue lines) using the algorithm proposed by \citet{Xiao2024}. The green and orange lines represent the contributions from RB and 4U1728-34, respectively.
It is clear that regular and repeating type-II bursts originate from RB. 
One type-I burst from 4U1728-34 (upper panel) was successfully distinguished.
}
\label{fig:persistent}
\end{figure}

We note that there is another persistent and luminous source 4U 1728-34  (also known as the Slow Burster; SB), close to our target with an angular separation of $\Delta \theta \approx$ 0.5$^\circ$.
Because the field of view (FoV) of {\it Insight}-HXMT/LE is larger than $\sim$1$^\circ$, these two sources cannot be distinguished straightforwardly.
In practice, the on-axis source was 4U 1728-34 for the 2017 observations, and type-II bursts emitted from RB were detected serendipitously \citep{Chen2021}.
Fortunately, the LE consists of collimators with three different orientations and FoVs (see Figure~\ref{fig:FOV}).
Two sources located at different positions within each FoV are associated with different effective areas.
This allows us to reconstruct the temporal information for each source when one of them is significantly variable, such as the bursting behavior that we considered  (for details, see the Appendix).
As examples, we show representative reconstructions in Figure~\ref{fig:persistent}, which indicates that the observed lightcurves can be successfully decomposed into contributions from different sources.
Meanwhile, the origin of each burst can be unambiguously identified. 
For the non-burst (persistent) emission, however, it was dominated by 4U1728-34. 
As a result, the systematic error of the decomposition algorithm prevents us from accurately estimating RB's persistent emission.

\section{Burst identification and classification}
\label{Burst_identification}
We first extracted background-subtracted lightcurves in the energy range of 2-10\,keV with a bin size of 1\,s,  where the background was estimated by using the tool \texttt{lebkgmap}.
We searched for bursts according to the criterion that the
burst peak flux is significantly larger than the average persistent emission at a confidence level of 5$\sigma$.
The persistent emission was calculated as follows.
For each GTI, we first calculated the mean count rate of all data points, and  then flagged points with count rates exceeding this value at a $5\sigma$ confidence level as burst candidates.
We excluded all flagged burst points, and repeated the above steps until all burst-related data points had been removed.
The average count rate of the remaining points was defined as the persistent level.
We note that since the bin size of lightcurves we used was 1\,second, sub-second bursts might be missed.
We defined the burst starting ($T_{\rm start}$) and ending ($T_{\rm end}$) times as the first and last points when the lightcurve reached the persistent level within 1$\sigma$ error.
In this study, we focused only on type-II bursts and excluded type-I bursts that can be identified according to their typical fast-rising and exponential decay profiles.
Actually, five type-I bursts were found in the dataset, and all of them originated from 4U 1728-34 instead of the Rapid Burster.
In total, 299 type-II bursts were identified in 2017, including 297 bursts that exhibited complete evolutionary profiles (i.e., with both rising and decay phases) and two additional events that were incomplete due to interrupted observations.
Using the same method, we observed 461 type-II bursts in 2020, including 13 incomplete bursts.
In 2017, only regular short bursts (known as mode-2) were detected, whereas in 2020, some long-duration bursts following a series of shorter bursts (mode-1) appeared in Epoch E and F (see insets of Figure \ref{fig:BAT}).
As mentioned above, the persistent emission cannot be accurately determined.
This prevents us from distinguishing between long and short bursts based on the presence of the ``hump" feature.
In practice, we defined bursts that occurred at the end of a burst sequence and had a longer duration as long bursts.


\section{Statistical properties of type-II bursts} 
\label{Results}
We characterized the morphology of the bursts quantitatively by defining a set of parameters in terms of their luminous and temporal properties: peak flux ($F_{\rm peak}$), fluence ($E$), burst duration ($t_{\rm dur}$), and recurrence time ($\Delta t$).
In this paper, the burst intensity is described by the count rate\footnote{For the sake of comparison, we converted count rates of 2017 bursts to the values that would be obtained for on-axis observations, i.e., multiplying by a factor that is the effective area ratio between off-axis and on-axis cases.}
in the LE energy band (2-10\,keV).
According to the Comptonization model reported by \citet{Chen2021}, one Eddington luminosity ($2.8 \times 10^{-8}\,{\rm erg\,s^{-1}\,cm^{-2}}$, see \citet{2015.T.Bagnoli} for assumptions) corresponds to 298\,counts/s.
We had also tested other models reported by \citet{2017van}, \citet{2004Falanga} and \citet{2000Masetti}, and found that changes of the flux-to-rate ratio were in general smaller than 15\%.
The peak flux is defined as the maximum count rate during bursts after subtracting the persistent emission.
The fluence is the integration of the burst count rate over their durations (with the persistent emission removed).
The recurrence time was defined as the time interval between the starting point of one burst and the subsequent burst, considering only consecutive bursts in uninterrupted GTIs and ignoring samples with gaps between bursts.
\label{statistical}

\subsection{Distributions of burst parameters}
\begin{figure}[htbp]
\begin{minipage}{1.0\linewidth}
\centerline{\includegraphics[width=\textwidth]{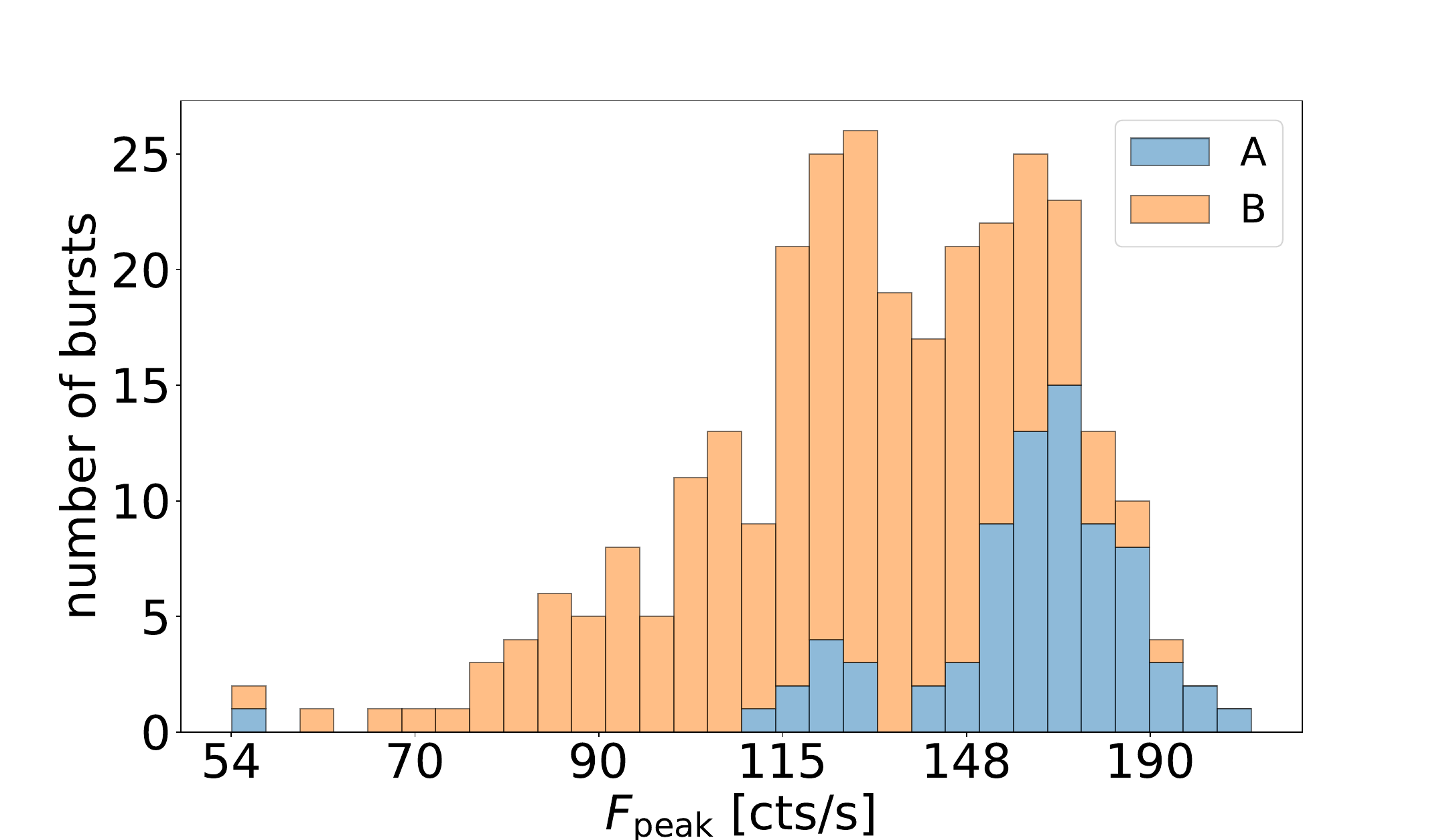}}
\centerline{\includegraphics[width=\textwidth]{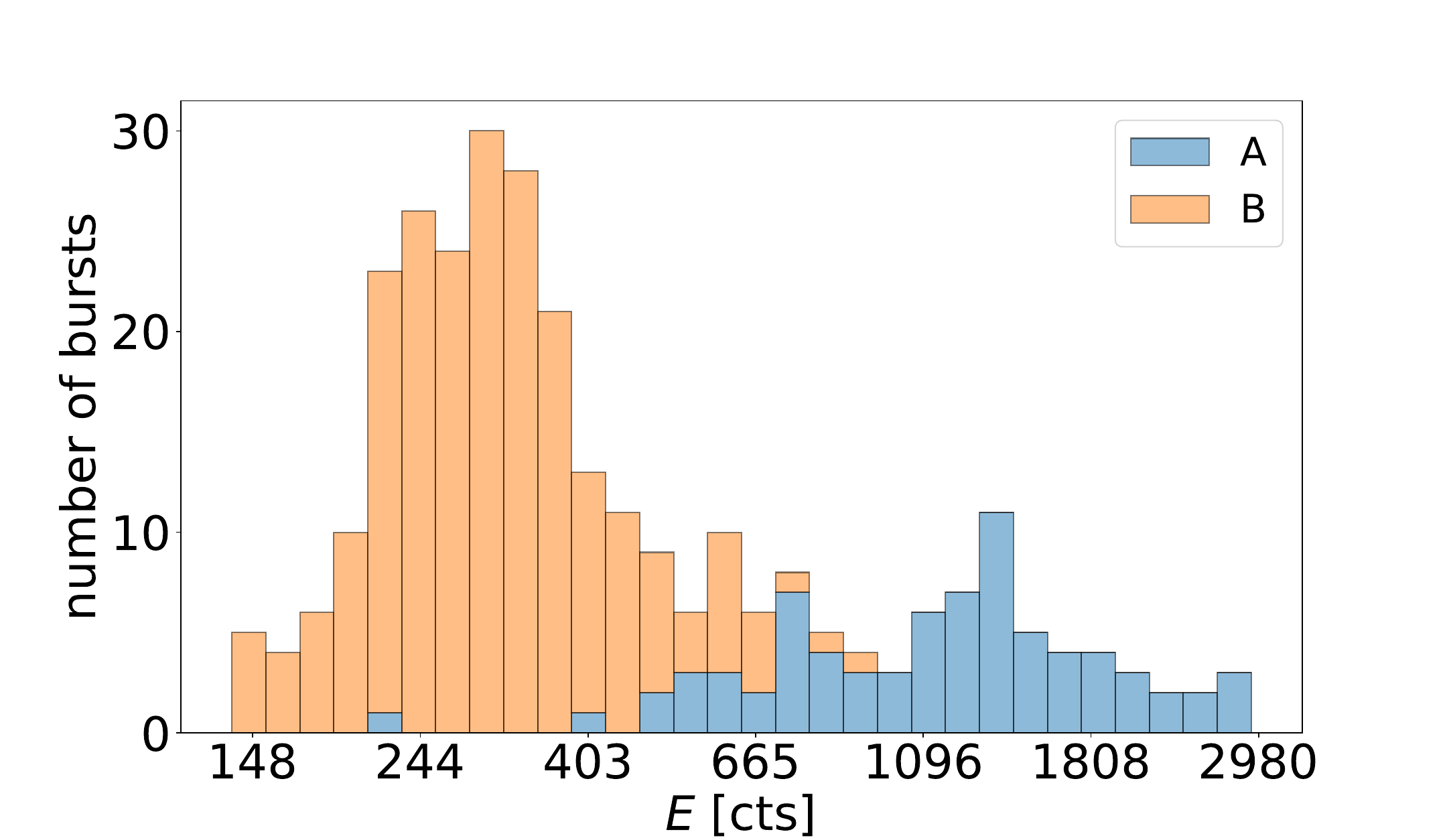}}
\centerline{\includegraphics[width=\textwidth]{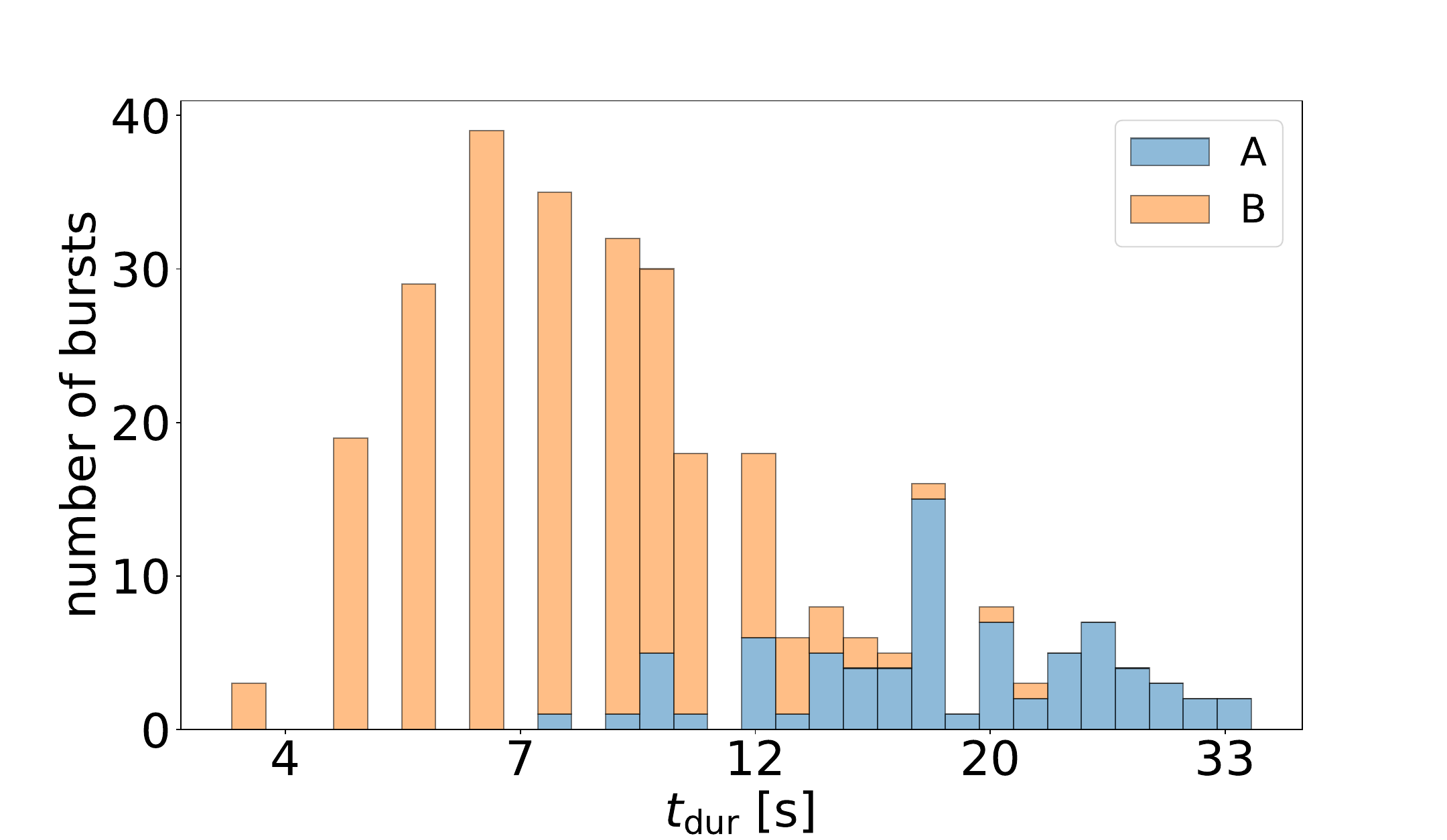}}
\centerline{\includegraphics[width=\textwidth]{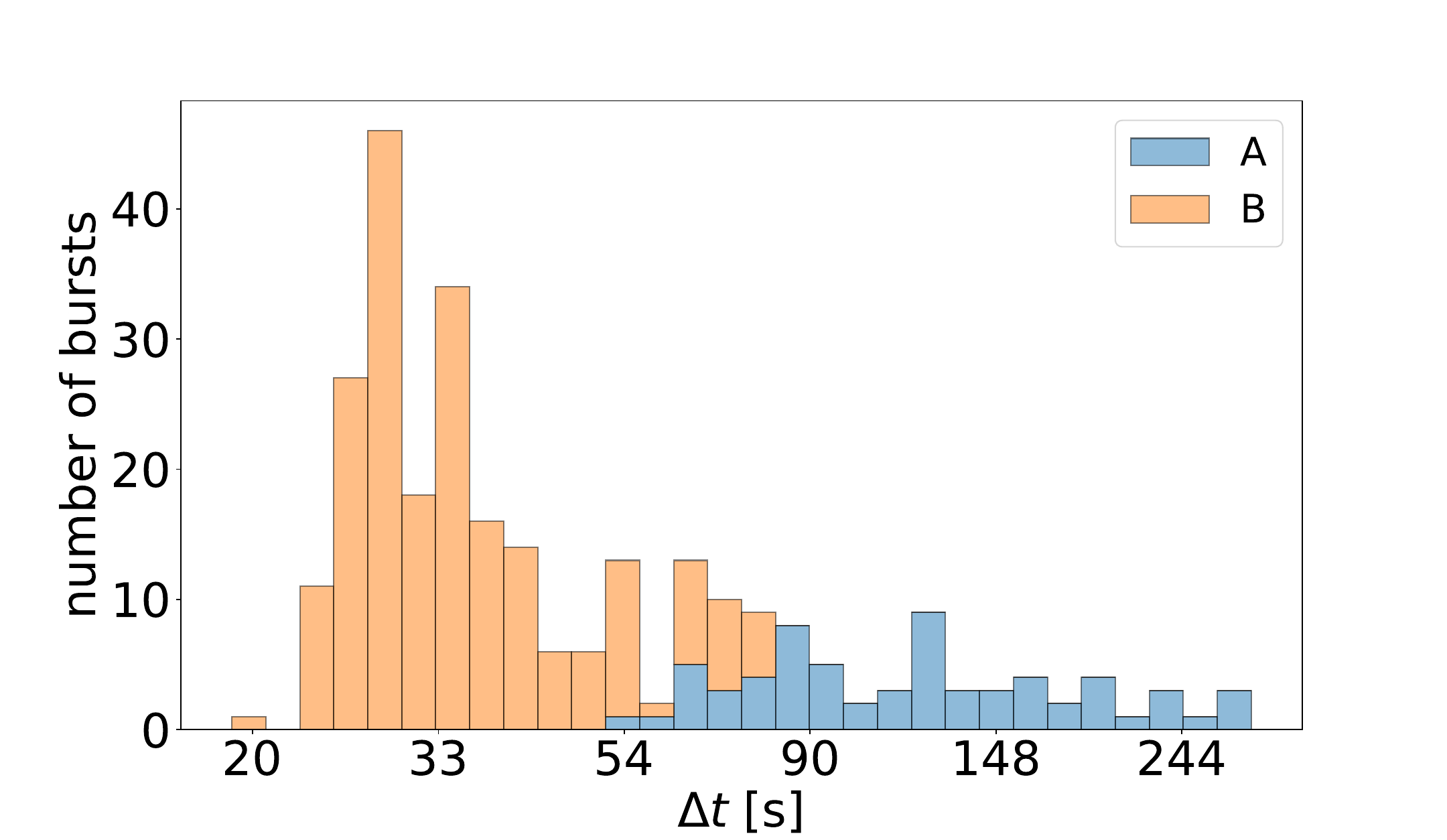}}
\centering
\end{minipage}
\caption{
Distributions of burst properties for bursts observed in 2017.
From top to bottom, the panels show the distributions of burst peak flux $F_{\rm peak}$, fluence $E$, duration $t_{\rm dur}$ and recurrence time $\Delta t$, respectively.
Each bar represents the total burst number, whereas different colors indicate the contributions of different  epochs.
In 2017 only short bursts were found.
}

\label{fig:properties2017}
\end{figure}

\begin{figure}[htbp]
\begin{minipage}{1.0\linewidth}
\vspace{3pt}
\centerline{\includegraphics[width=\textwidth]{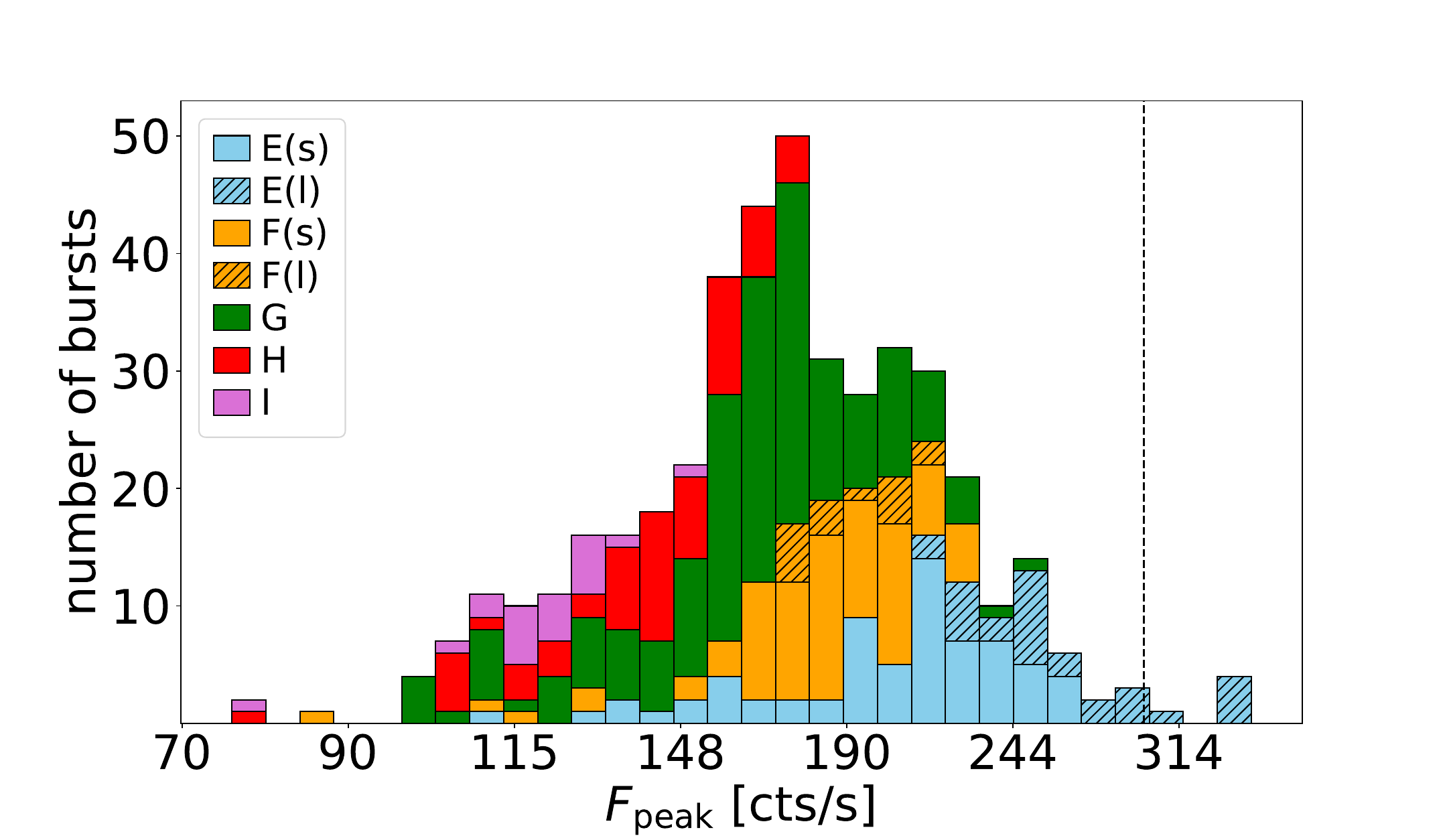}}
\vspace{3pt}
\centerline{\includegraphics[width=\textwidth]{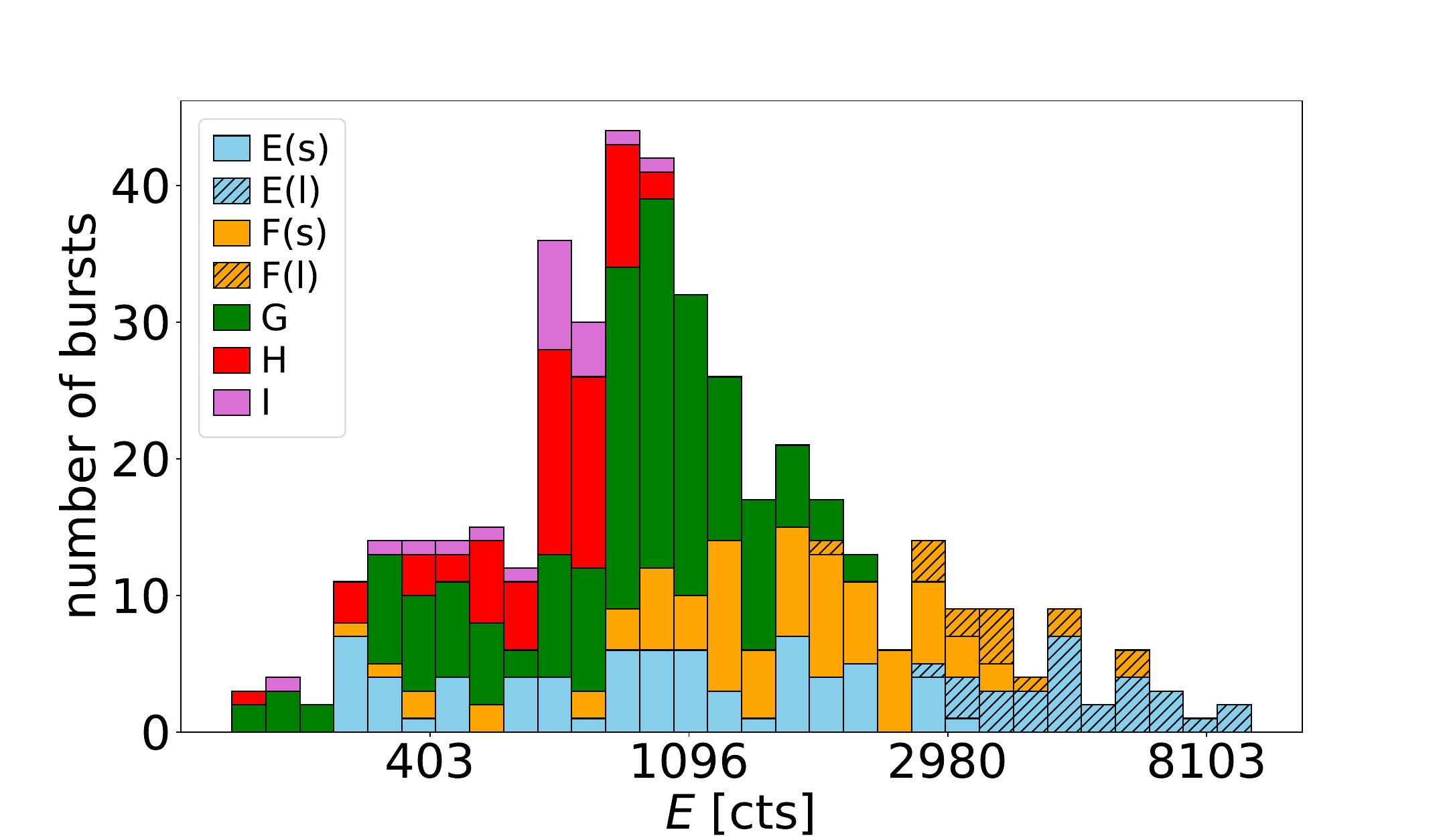}}
\centerline{\includegraphics[width=\textwidth]{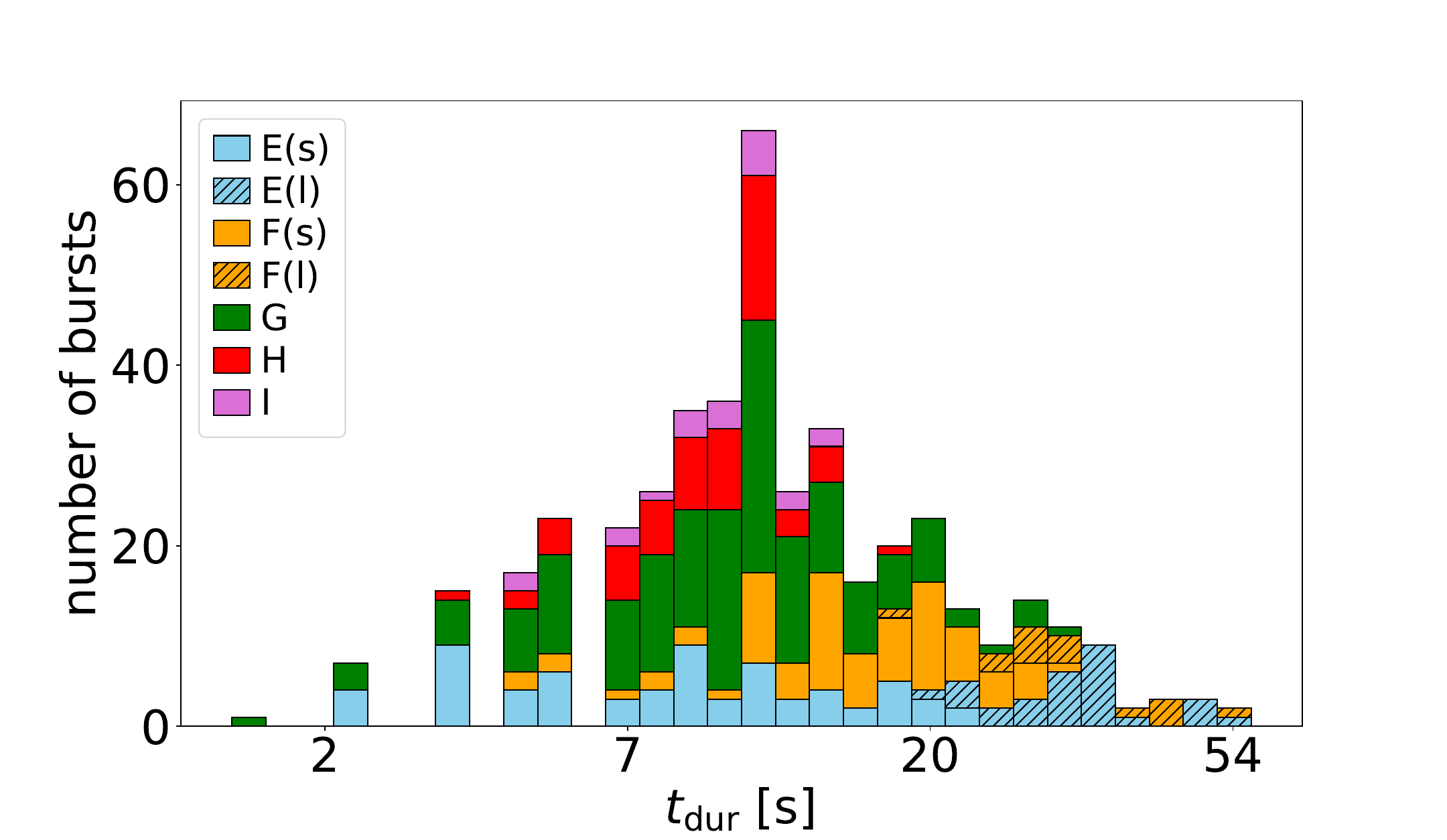}}
\centerline{\includegraphics[width=\textwidth]{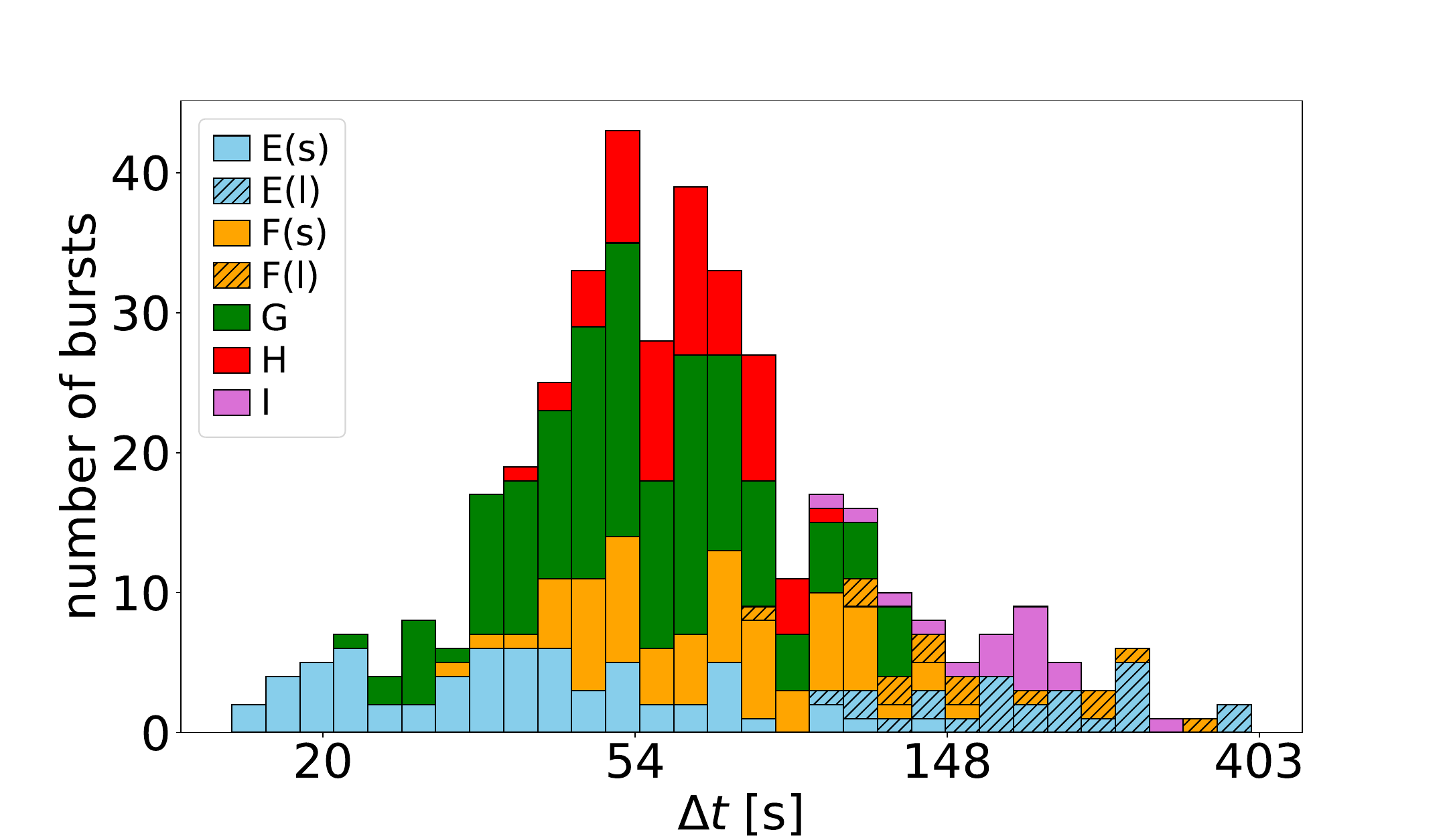}}
\centering
\end{minipage}
\caption{
Distributions of burst properties for bursts observed in 2020.
From top to bottom, the panels show the distributions of burst peak flux $F_{\rm peak}$, fluence $E$, duration $t_{\rm dur}$ and recurrence time $\Delta t$, respectively.
The black dashed line in the top panel represents the Eddington limit. 
The regions shaded with right-diagonal hatching correspond to long bursts.
The label "(s)" and "(l)" refer to short and long bursts if both of them present in the same epoch. 
}
\label{fig:properties2020}
\end{figure}

In Figures~\ref{fig:properties2017} and \ref{fig:properties2020}, we present the distributions of burst properties for events detected in 2017 (Epochs A, B) and 2020 (Epochs E–I), respectively.
Because the burst morphology changes along with the outburst evolution, we show them as separate samples with different colors.
For bursts in 2017, the peak flux distribution in Epoch B is centered around 120\,cts/s, significantly lower than those in Epoch A.
For 2020 bursts, their peak fluxes are comparable to those of 2017 bursts, and present a consistent decreasing trend over time.
We find that peak fluxes of nearly all bursts are below the Eddington limit (the vertical line in the top panel of Figure~\ref{fig:properties2020}), with the exception of several long bursts in Epoch E.

The burst fluence distributions also show clear differences between epochs. 
For example, the distribution for Epoch B peaks at~300\,cts, while for Epoch A the fluence spans primarily from 400\,cts to 3000\,cts (the middle panel of Figure~\ref{fig:properties2017}). 
For 2020 bursts, those in Epochs E and F exhibit a much broader fluence distribution than others, because they occurred in mode-1 which have more diverse morphologies.

The burst duration distributions are shown in the third panels of Figures~\ref{fig:properties2017} and \ref{fig:properties2020}. 
For the 2017 data, all burst durations in Epochs A and B are shorter than 35\,s, with a large fraction of Epoch B bursts lasting less than 6\,s.
For the 2020 data, burst durations in Epochs E and F (mode 1) span from a few seconds to over 50\,s, while durations in Epochs G–I (mode 2) are centered around 12\,s.

The burst recurrence time distributions are shown in bottom panels of Figures~\ref{fig:properties2017} and \ref{fig:properties2020}.
In 2017, recurrence times range from 20\,s to 250\,s, with samples in Epoch B significantly lower than those in Epoch A. 
In 2020, recurrence times span a wider range for those in Epochs E and F.

In summary, we find that from Epoch A to B bursts become shorter in duration and less energetic but more frequent.
For bursts in 2020, those in Epochs E and F (mode 1) show significant diversity in all parameters except the peak flux, while burst properties in Epochs G–I (mode 2) are consistent with those observed in Epochs A and B.

\subsection{Correlations between burst parameters}

\begin{figure}[htbp]
\centering
\includegraphics[width=0.5\textwidth]{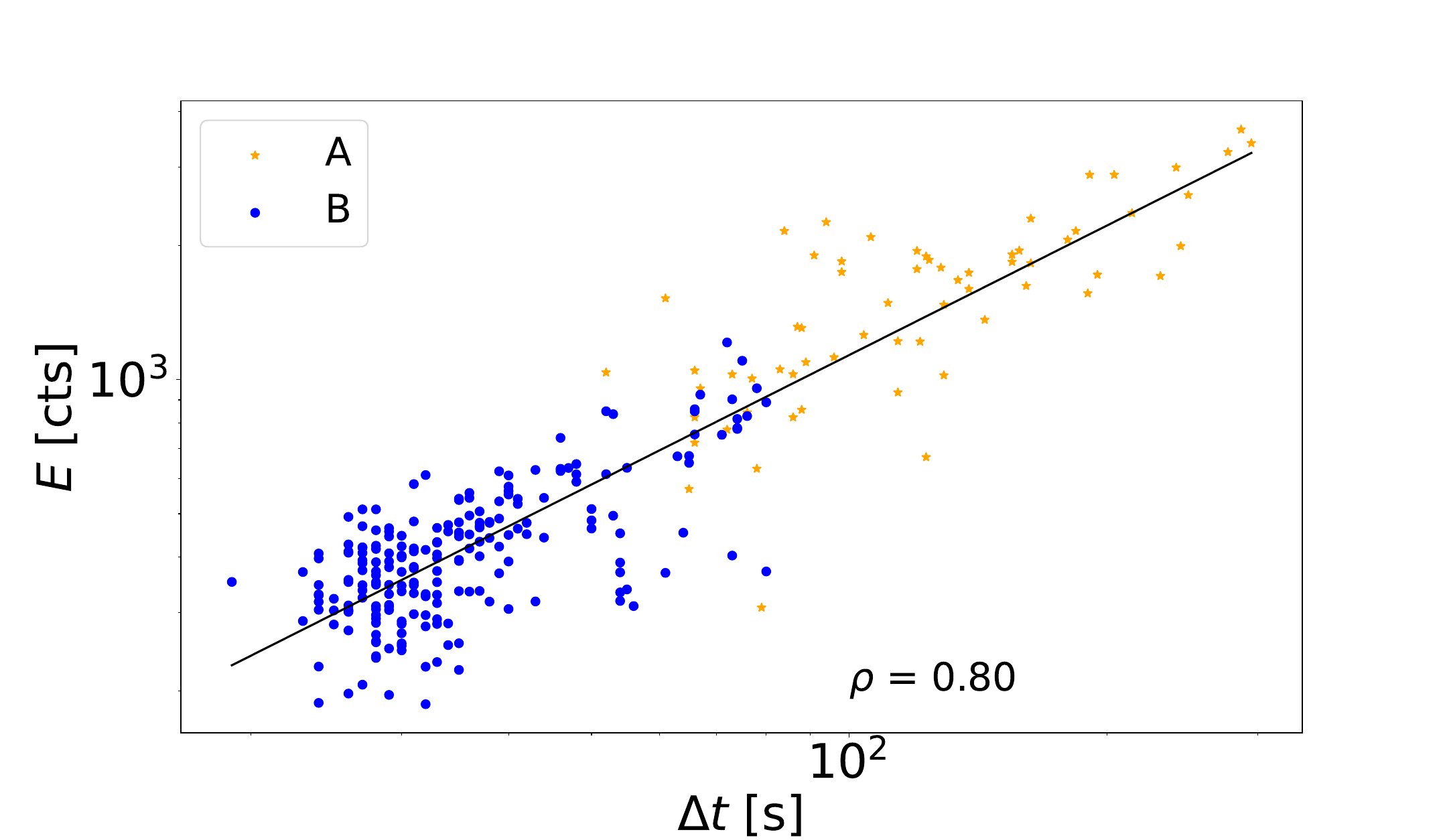}
\includegraphics[width=0.5\textwidth]{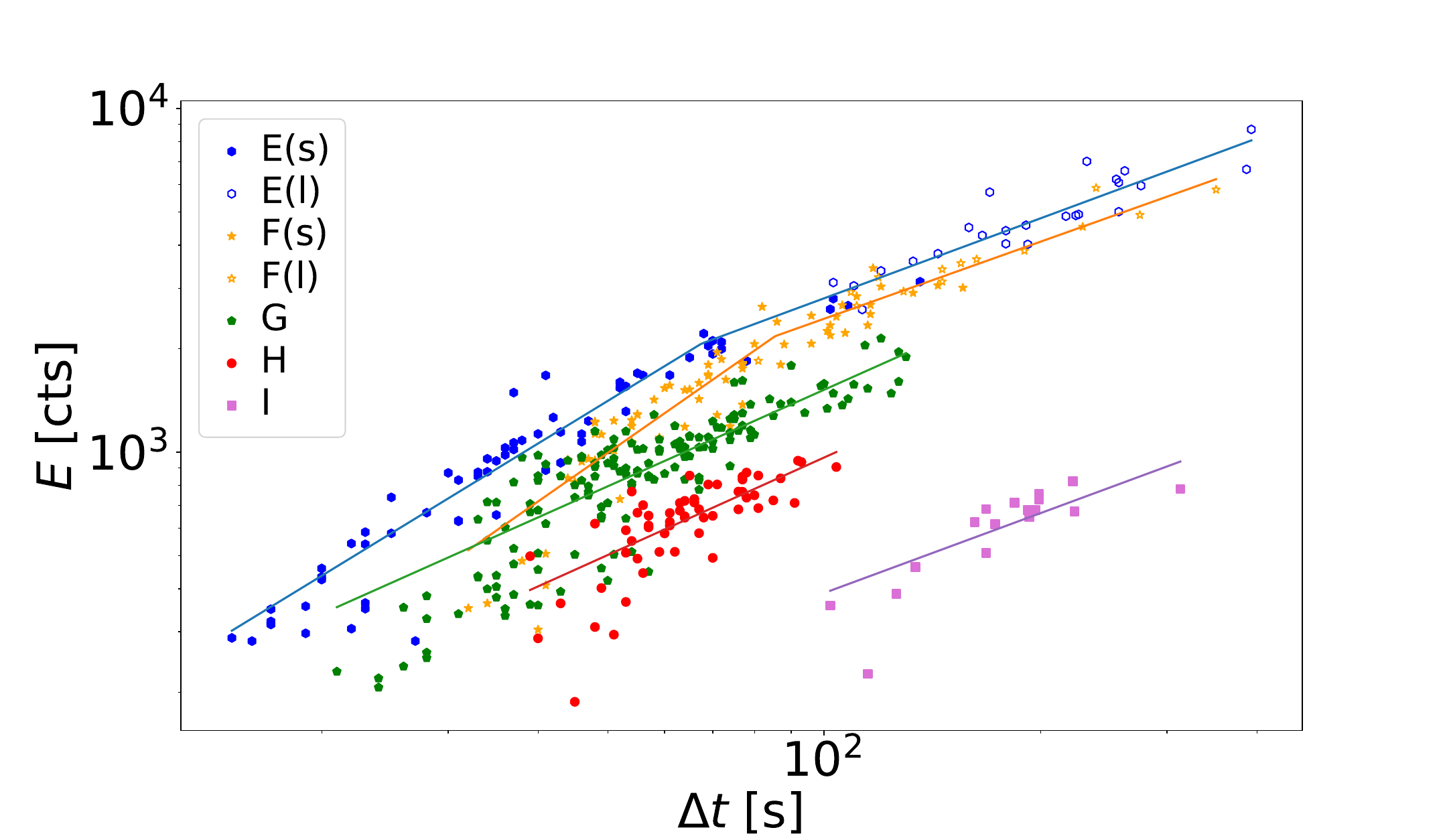}
\caption{The relationship between the fluence ($E$) and the recurrence time ($\Delta t$) for bursts detected in 2017 (upper) and 2020 (bottom), where different colors present different epochs.
Open and filled markers denote long and short bursts. A linear relation is evident in 2017, while a broken powerlaw (BPL) model is used for Epochs E and F in 2020 (best-fit parameters in Table \ref{tab:relaxation}).
In the upper panel we also show the Spearman correlation coefficient $\rho$.}
    \label{fig:relaxation}
\end{figure}

In this subsection, we investigate the correlations between the aforementioned burst parameters.
Figure~\ref{fig:relaxation} depicts the relaxation oscillator behavior, i.e., the relation between burst fluence $E$ and recurrence time $\Delta t$. 
For 2017 bursts, a linear $E-\Delta t$ relation is clearly found with a Spearman correlation coefficient $\rho=0.80$.
This relation can be described as $E \propto \Delta t^{\alpha}$ where the index ${\alpha}={ 0.95 \pm 0.02}$ was determined using the least-square method. 
In practice, we defined $\chi^2 = \sum_i \frac{(D_{\rm i} - M_{\rm i})^2}{\sigma_{\rm sys}^2}$, where $D_{\rm i}$ is the observed value, $M_{\rm i}$ is the value predicted by the model.
The systematic error $\sigma_{\rm sys}$ was included to account for the systematic scattering, and was estimated under the assumption that reduced-$\chi^2$ is equal to 1 for the best-fitting model.
With this systematic error, we calculated the uncertainties of parameters as $\Delta\chi^2=1$, corresponding to a confidence level of 68\%.
For the 2020 bursts, such a relation is not significant when considering all bursts as a sample.
However, it can be revealed if we consider bursts in separate epochs.
Clear downward offsets can be seen with the evolution of the outburst when comparing $E$–$\Delta t$ relations obtained from different epochs.
These offsets are indeed expected because observations at different accretion rates would accumulate the same mass at different time scales to trigger bursts having comparable $E$. 
In addition, we found that even within the same epoch the index $\alpha$ seems to be not a constant.
To quantify this possibility, we compared two models for fitting the $E$–$\Delta t$ relation: 1) the aforementioned powerlaw model, and 2) a broken powerlaw model (BPL) which describes as 
\[{E=}\left\{ \begin{array}{l}
K \Delta t^{\alpha_1}, when\ \Delta t<t_{\rm b}\\
K \Delta t^{\alpha_2}, when\ \Delta t\geq t_{\rm b}
\end{array} \right.\]
We determined the $\sigma_{\rm sys}$ using the BPL model, and then used this value to fit both two models.
We compared their goodness-of-fits using F-test, and found that in Epochs E and F the BPL model improves the fits at a significance level of $>3\sigma$.
In addition, the index $\alpha_2$ in the high $\Delta t$ range is much smaller than $\alpha_1$ (see Table~\ref{tab:relaxation}).
We consider that this might reflect the intrinsic difference between long and short bursts, given that the critical value $t_{\rm b}$ approximately separates these two groups.
For Epochs G-I, the improvement is not significant, so we still keep the powerlaw model.
We note that although there are offsets in Figure~\ref{fig:relaxation} across Epochs G-I, no evident variation in $\alpha$ was found (see Table~\ref{tab:relaxation}).

\begin{figure}
\centering
\includegraphics[width=1\linewidth]
{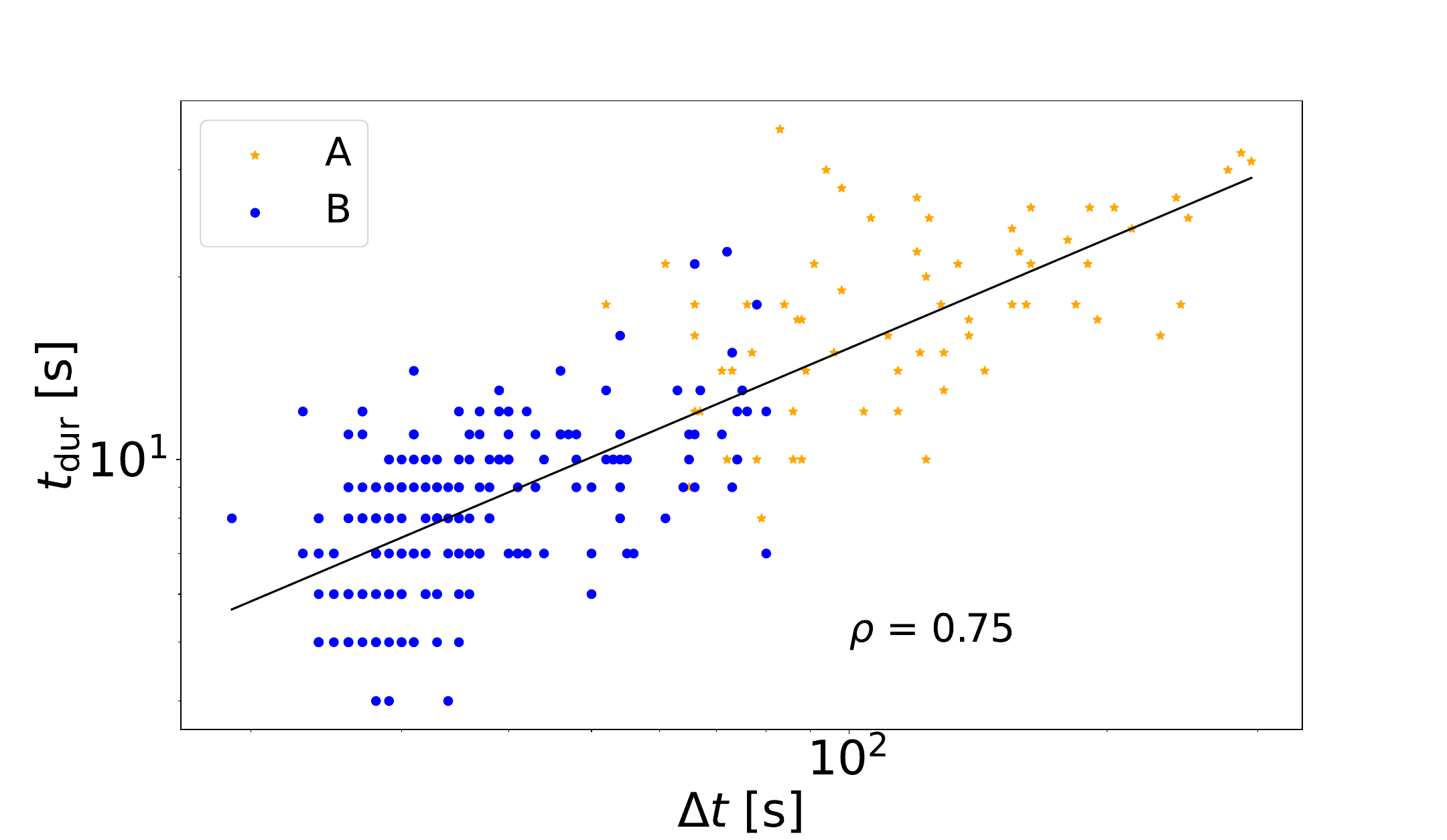}
\includegraphics[width=1\linewidth]
{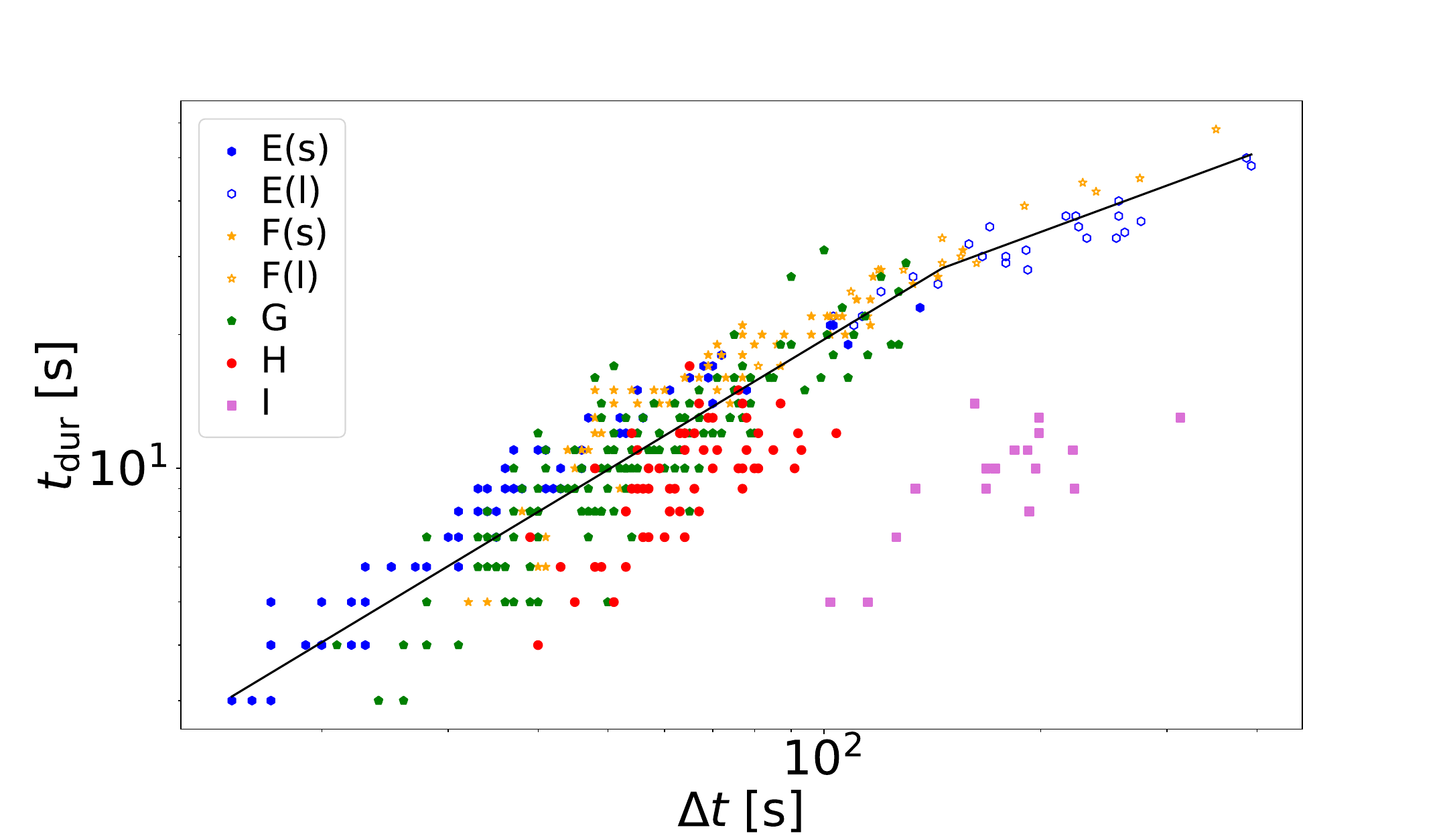}
\caption{The relationship between the duration $t_{\rm dur}$ and the recurrence time $\Delta t$ for 2017 (upper) and 2020 (lower) bursts.
Black lines show the best-fit results with a powerlaw model (upper) and a BPL model (bottom).
Bursts in Epoch I are excluded from the analysis.
In the upper panel we also show the Spearman correlation coefficient $\rho$.}
\label{fig:duration-recur}
\end{figure}

As suggested by \citet{2015.T.Bagnoli}, the burst duration $t_{\rm dur}$ is a better parameter than the fluence for the $E-\Delta t$ relation.
We further investigated the relationship between $t_{\rm dur}$ and the recurrence time $\Delta t$ (see Figure~\ref{fig:duration-recur}).
For 2017 bursts, we found $t_{\rm dur} \propto \Delta t^{ 0.57 \pm 0.02}$
with a Spearman correlation coefficient $\rho = 0.75$. 
Probably due to the small sample, this correlation coefficient is not larger than that of the $E-\Delta t$ relation.
For 2020 bursts, on the contrary, the $t_{\rm dur}-\Delta t$ relation of bursts is indeed more significant than the $E-\Delta t$ relation across different Epochs. 
If excluding outliers from Epoch I, similar to the $E-\Delta t$ relation, the $t_{\rm dur}-\Delta t$ relation can also be fitted with a BPL model, which is superior to a powerlaw model at a confidence level of {$5\sigma$}.
The resulting indices are $0.98 \pm 0.03$ and $0.60\pm0.05$ when $\Delta t$ is less than or greater than the break point $\Delta t=146\pm14$\,s.

\begin{figure}[htbp]
\centering
\includegraphics[width=0.5\textwidth]{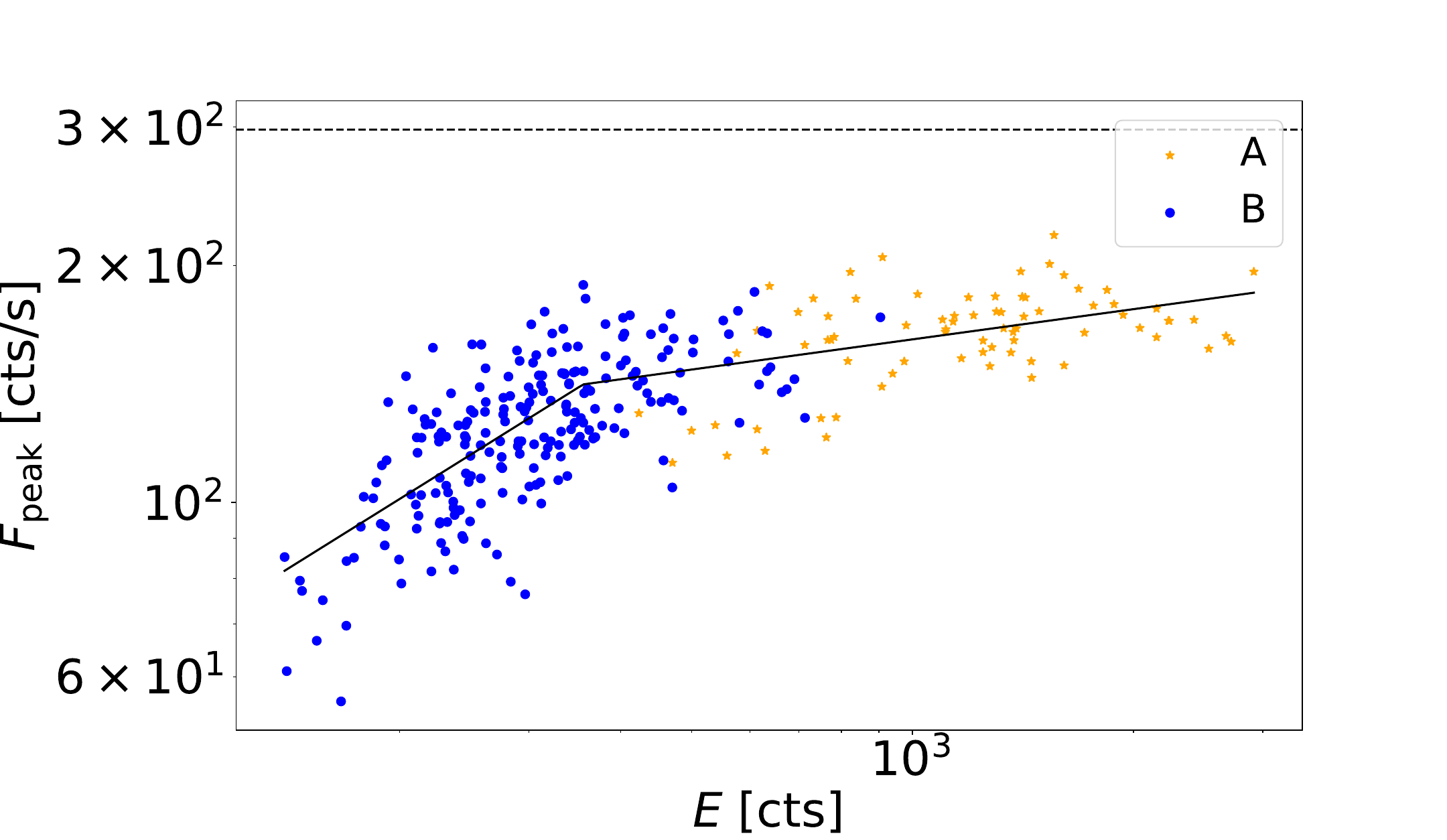}
\includegraphics[width=0.5\textwidth]{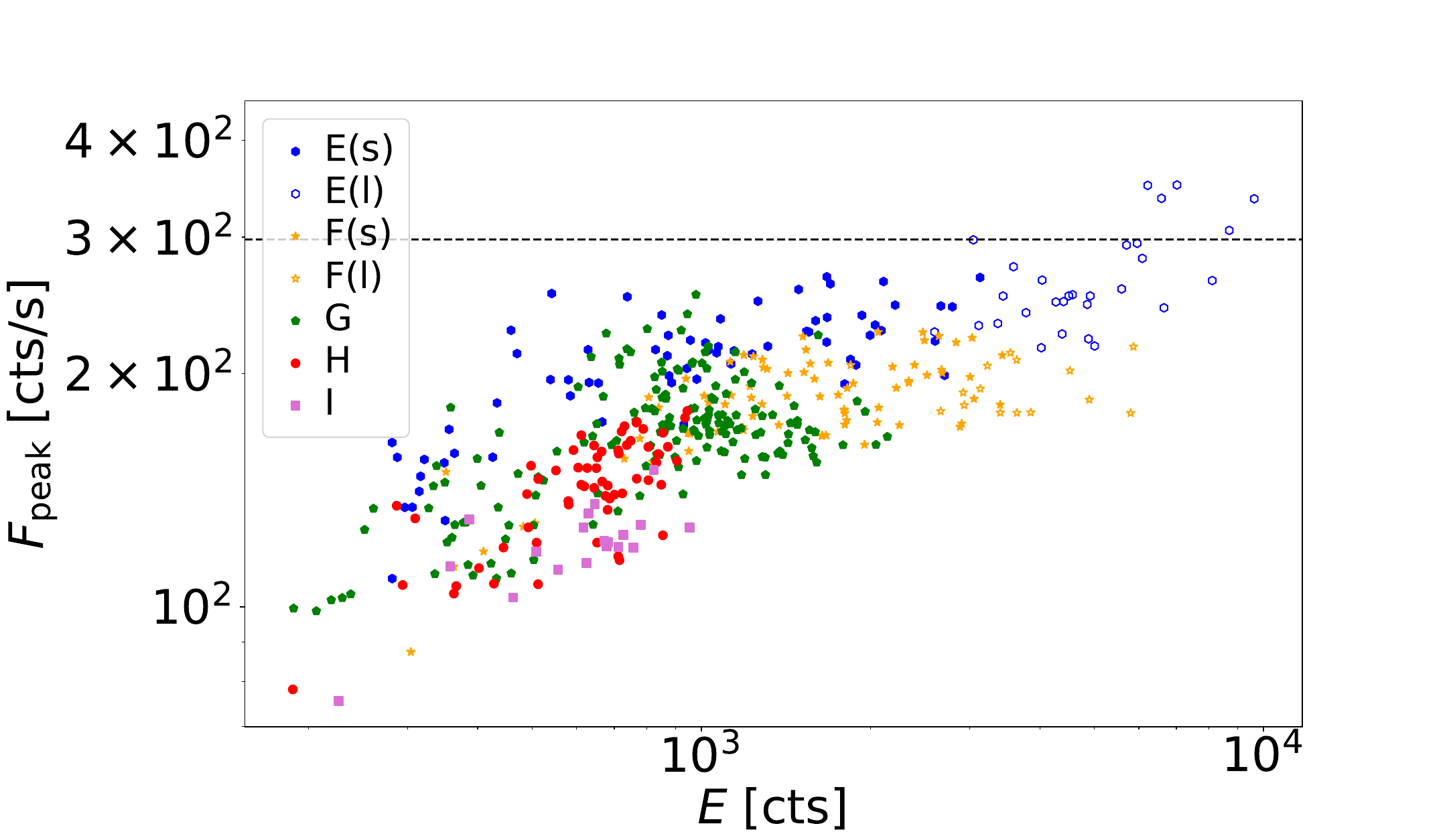}
\caption{The relationship between the peak flux $F_{\rm peak}$ and the fluence $E$ for bursts in 2017 (upper) and 2020 (bottom), where colors and and symbols are the same as those in Figure~\ref{fig:relaxation}.
The dashed line indicates the Eddington luminosity.
The black line in the top panel represents the best fitting BPL model.}

    \label{fig:peak-fluence}
\end{figure}

We also investigated correlation between the peak flux ($F_{\rm peak}$) and the fluence ($E$).
As shown in Figure~\ref{fig:peak-fluence} top panel, the burst peak flux in 2017 increases with the fluence in the low fluence range, while the increasing rate slows significantly once $F_{\rm peak}$ reaches  137 $\pm$ 9 cts/s.
This suggests a saturation behavior for the flux larger than $F_{\rm peak}$, which corresponds to approximately  46\% of the Eddington luminosity.
We also studied $F_{\rm peak}-E$ relations for 2020 bursts, and found that they could be described by a powerlaw model without the need for a more complex BPL model.

\begin{figure}[htbp]
\centering
\includegraphics[width=0.5\textwidth]{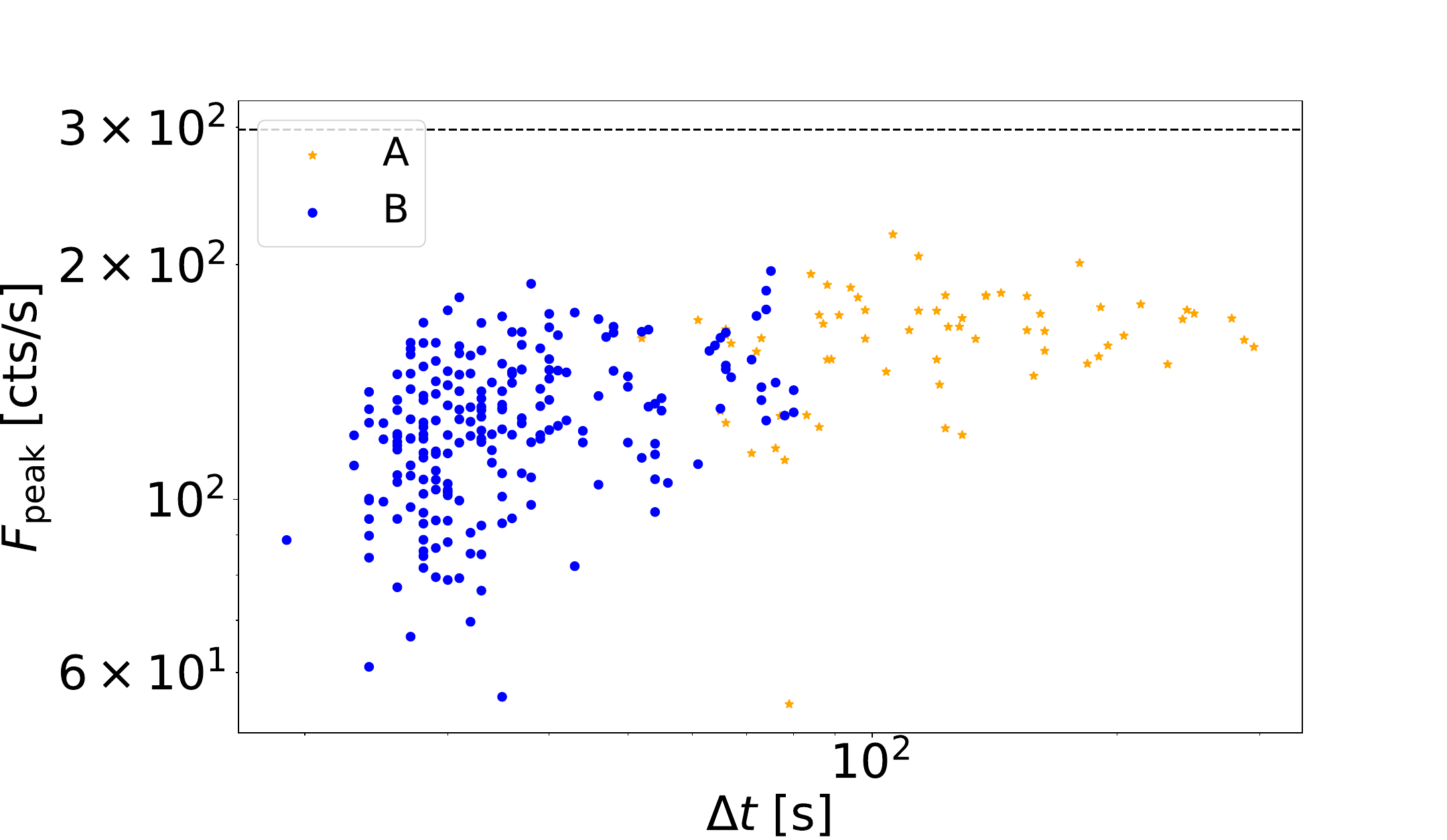}
\includegraphics[width=0.5\textwidth]{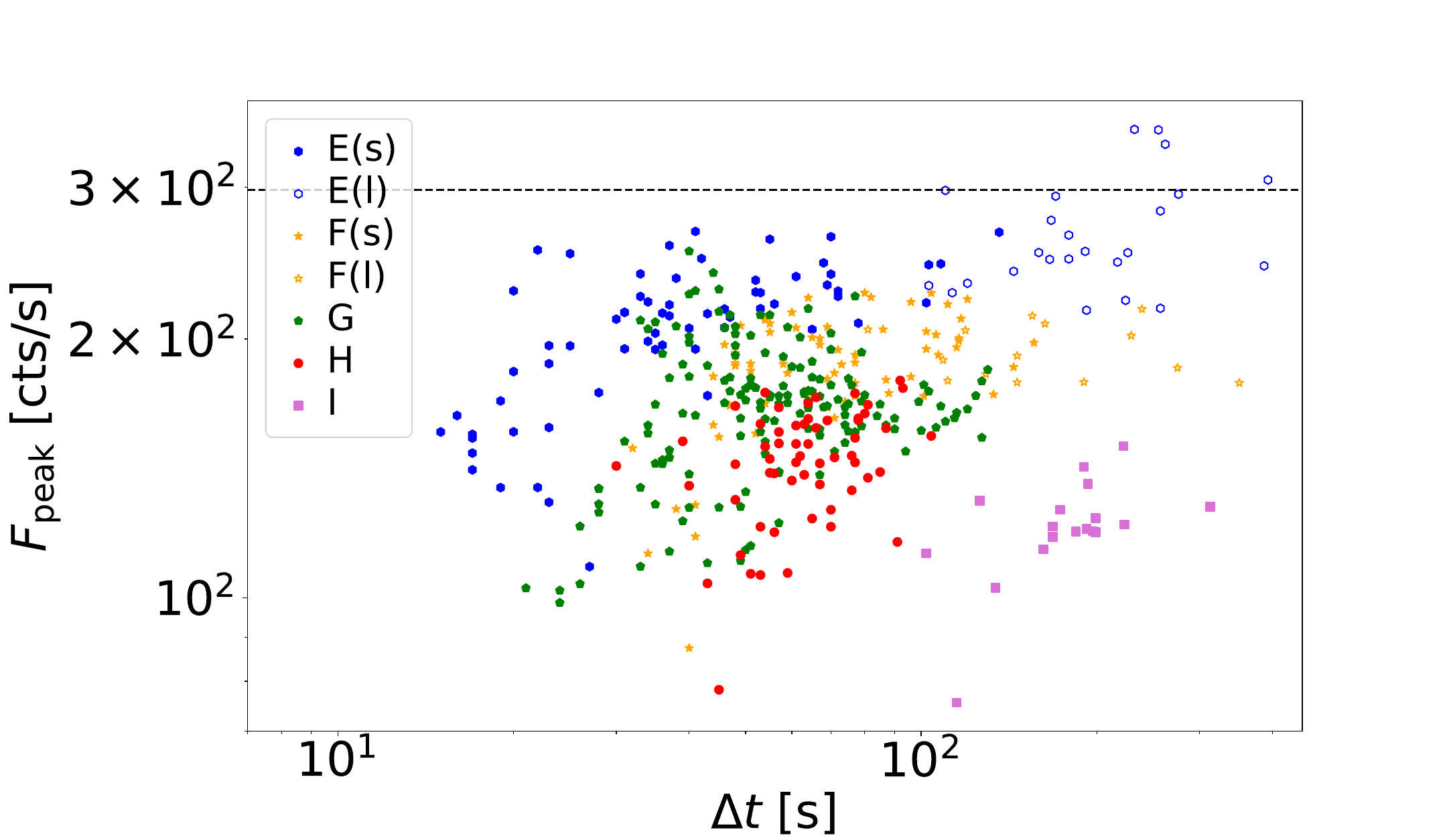}
\caption{The relationship between the peak flux $F_{\rm peak}$ and the recurrence time $\Delta t$ for bursts in 2017 (upper) and 2020 (bottom), where colors and and symbols are the same as those in Figure~\ref{fig:relaxation}. 
The dashed line indicates the Eddington luminosity.
    }
    \label{fig:peak-recurrence}
\end{figure}

After investigating the $E-\Delta t$ and $F_{\rm peak}-E$ relations, we further tested whether $F_{\rm peak}$ is correlated with $\Delta t$.
As presented in Figure~\ref{fig:peak-recurrence} top panel, bursts in 2017 might exhibit a weak $F_{\rm peak}-\Delta t$ correlation, but it is not as obvious as the $E-\Delta t$ relation.
For bursts in 2020 (bottom panel), no evident $F_{\rm peak}-\Delta t$ relation was found, neither across different epochs nor within individual epochs.

\begin{figure}
    \centering
    \includegraphics[width=1\linewidth]
    {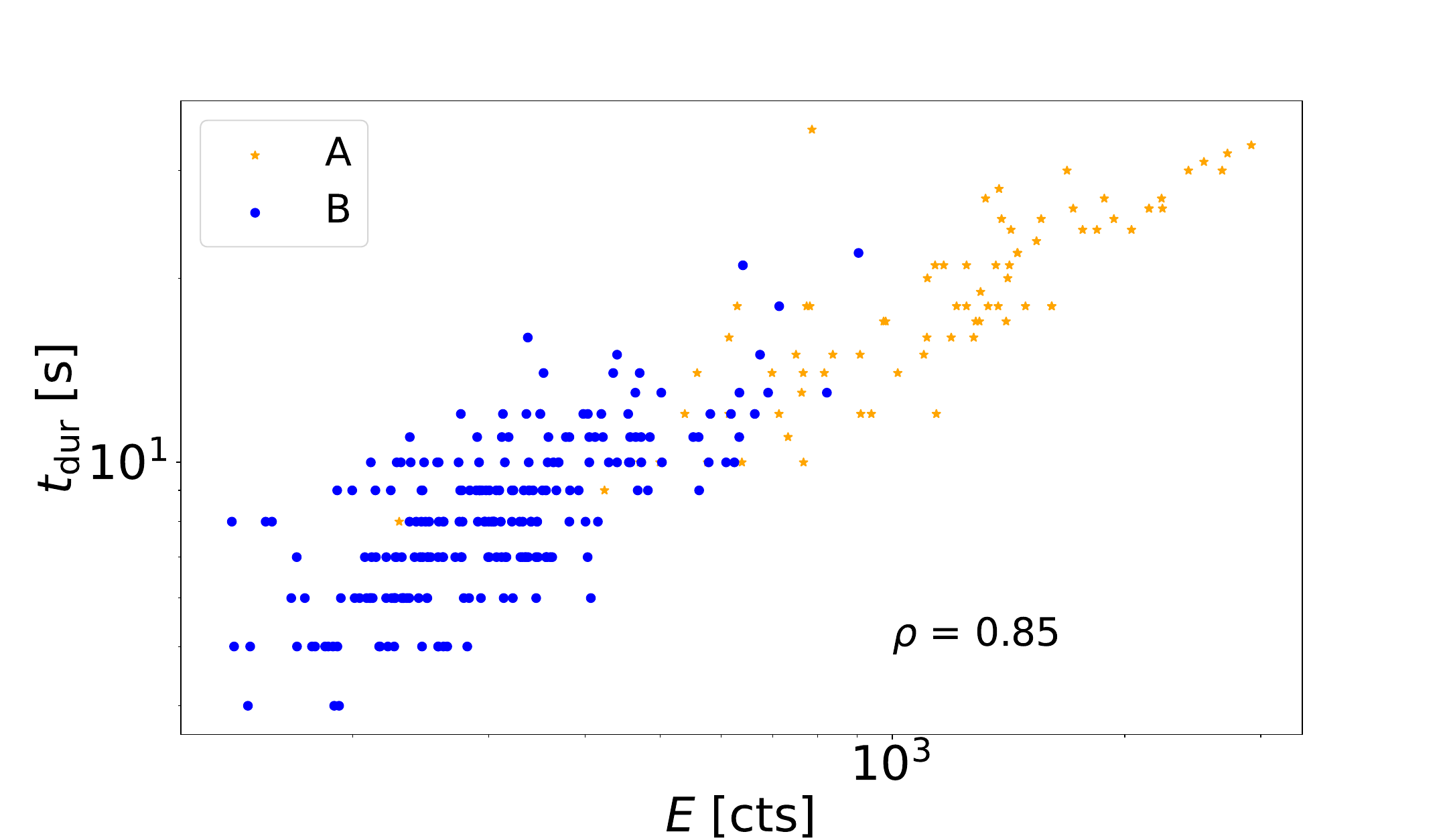}
    \includegraphics[width=1\linewidth]
    {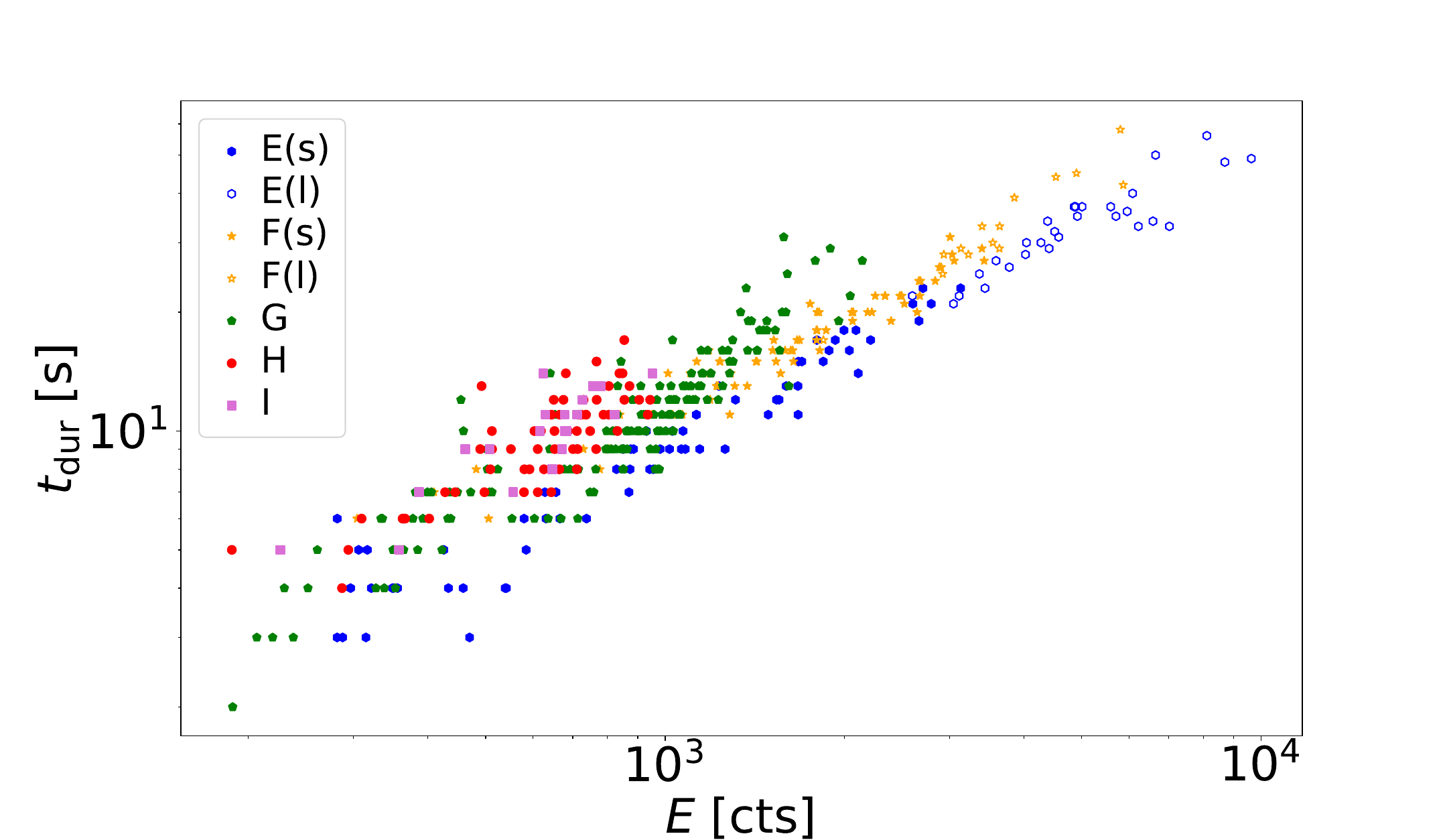}
    \caption{The relationship between the duration $t_{dur}$ and the fluence $E$ for 2017 (upper) and 2020 (lower) bursts, where colors and and symbols are the same as those in Figure~\ref{fig:relaxation}.
    A strong positive correlation is indicated by the Spearman correlation coefficient $\rho=0.85$ in the upper panel. A similar positive correlation is also evident across epochs for 2020 bursts.
    }
    \label{fig:duration-Eb}
\end{figure}

Based on the $E-\Delta t$ and $t_{\rm dur} - \Delta t$ relations, we further investigated the correlation between $t_{\rm dur}$ and $E$.
As shown in Figure~\ref{fig:duration-Eb}, a linear correlation is evident for both 2017 
and 2020 bursts.
Given the saturated peak flux discussed above  and the fact that $E$ is approximately as $F_{\rm peak}\times t_{\rm dur}$, it is not surprising that bursts with a larger fluence exhibit a longer duration.

\begin{table}[h]
    \centering
    \small
    \caption{Best fitting parameters of $E-\Delta t$ relations, as shown in Figure~\ref{fig:relaxation}, using a broken powerlaw model (Epochs E and F) or a powerlaw model (Epochs G-I).
    }
    \begin{tabular}{ccccc}
        \toprule
        Epochs & $\alpha$ & $\alpha_1$ & $\alpha_2$ & $t_{\rm b}$ (s)\\
        \hline
        E  & - & 1.28$\pm$0.17 & 0.77$\pm$0.03 & 67.51$\pm$13.56  \\
        F & - & 1.46$\pm$0.15 & 0.74$\pm$0.04 & 85.46$\pm$6.75  \\
        G & 0.93$\pm$0.04 & - & - & -\\
        H & 0.94$\pm$0.10 & - & -  & -\\
        I & 0.77$\pm$0.14 & - & -  & -\\
        \hline
    \end{tabular}
    \label{tab:relaxation}
\end{table}

\section{Discussion and Summary}
\label{Discussion}

\begin{figure}
    \centering
    \includegraphics[width=1\linewidth]
    {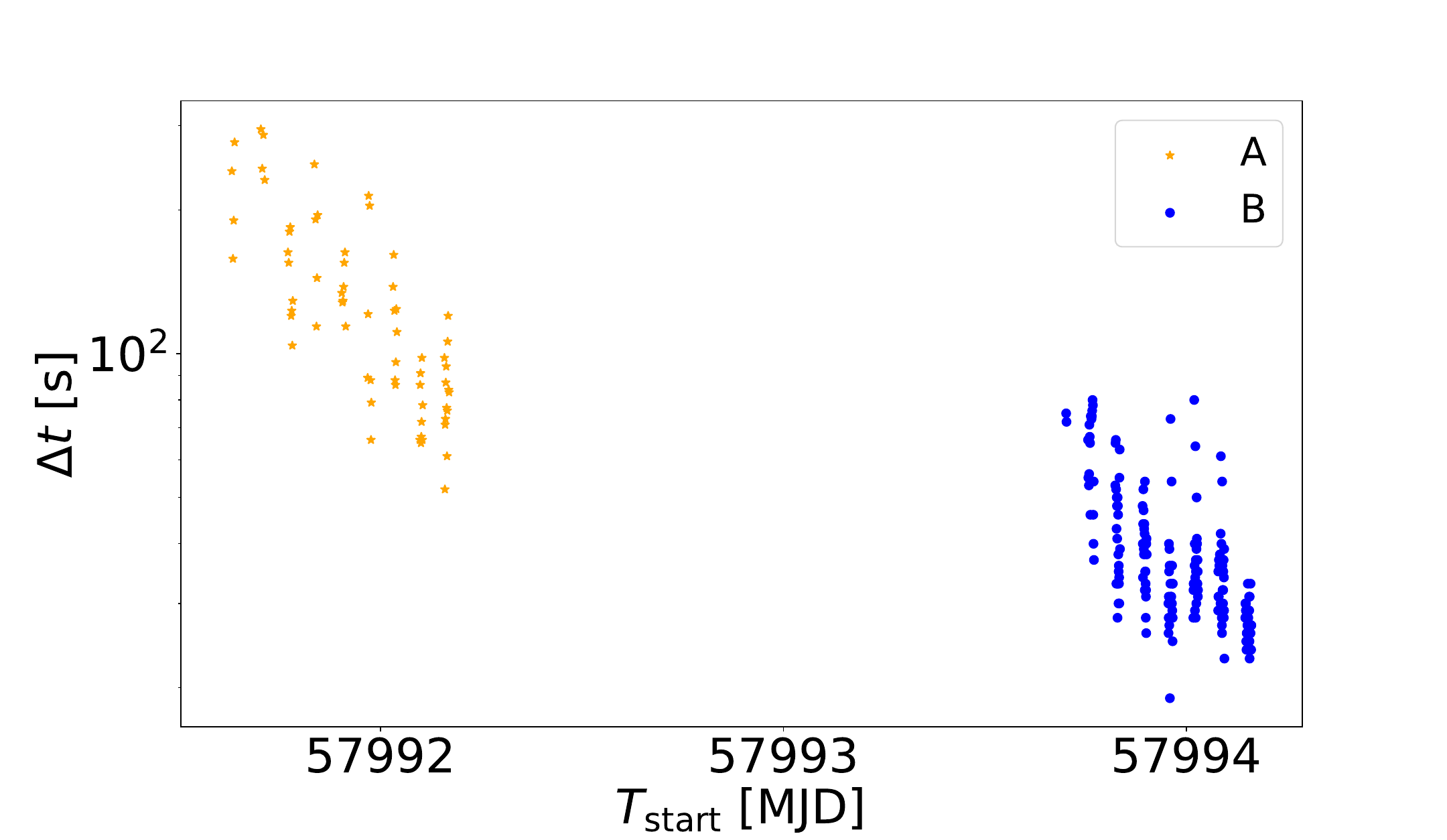}
    \includegraphics[width=1\linewidth]
    {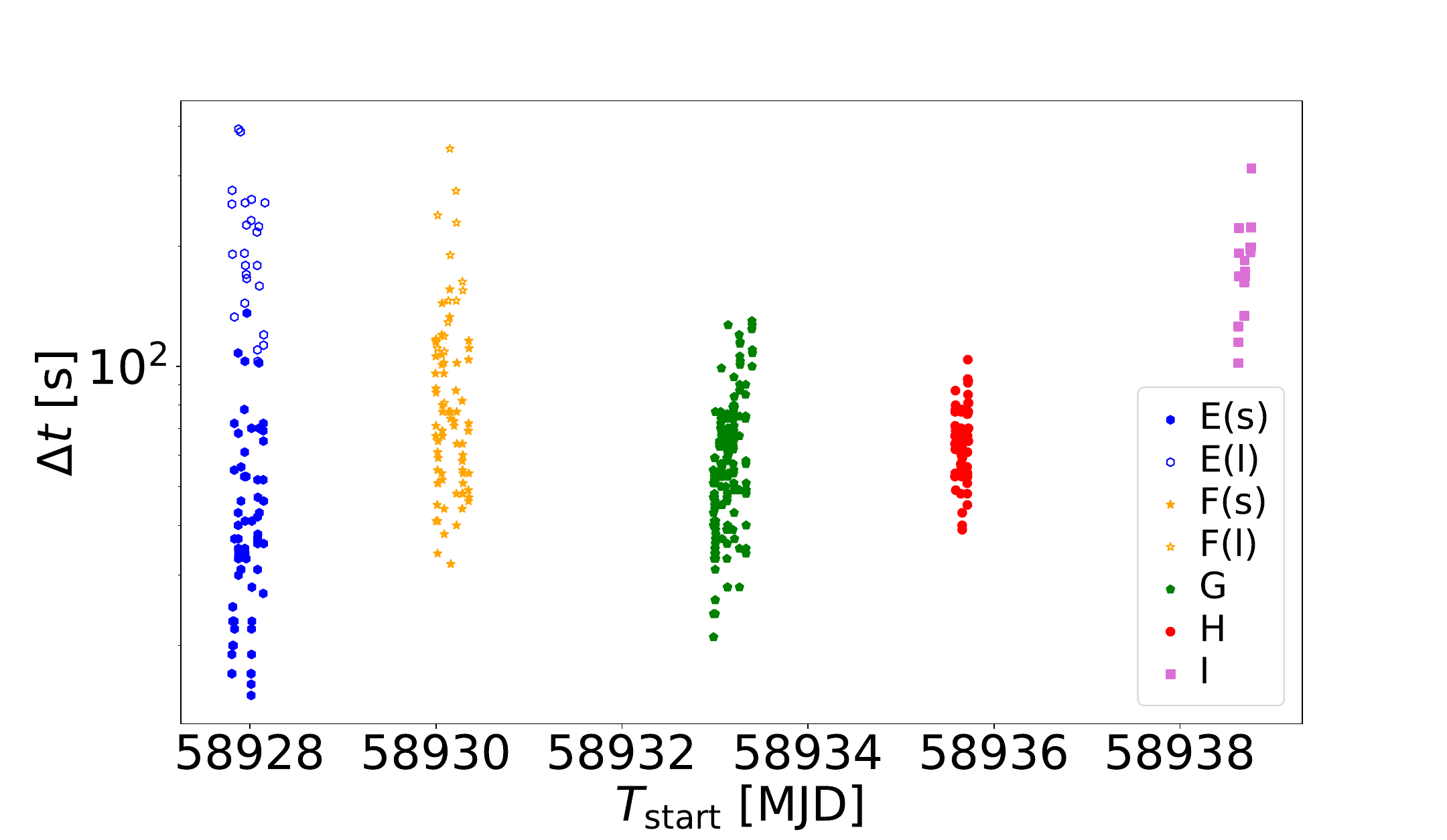}
    \caption{
    Evolution of burst recurrence time $\Delta t$ along with outbursts in 2017(upper) and 2020(lower), where colors and and symbols are the same as those in Figure~\ref{fig:relaxation}.
   We find a decreasing trend in 2017. In 2020, the averaged recurrence time for Epoch I bursts is significantly longer than those for other Epochs.}
   
    \label{fig:recur_start}
\end{figure}

We performed a thorough search for type-II bursts in Rapid Burster with {\it Insight}-HXMT data observed in 2017 and 2020.
Compared with the previous report by \citet{Chen2021}, we applied looser GTI selection criteria with an advanced identification method to optimize burst detections.
In total, we identified 299 and 461 bursts in the decay phases of the 2017 and 2020 outbursts, respectively.
Along with outbursts, these bursts exhibit significant variations in burst properties, such as the peak flux $F_{\rm peak}$, fluence $E$ and recurrence time $\Delta t$.
For instance, $\Delta t$ presents a wide range spanning from 10\,s to 400\,s.
In the literature, bursts can be divided into several types according to their morphologies, which usually appear in different outburst phases \citep[for a review, see][]{1993Lewin}.
For the 2020 observations, we detected the transition from non-burst epochs (C and D) to mode-1 bursts (E and F), and then switched to the mode-2 burst regime (G-I).
However, this evolutionary trend is not rigorous and may show back and forth changes.
For example, during the decay phase of an outburst, bursts are expected to become increasingly sparse.
For the 2017 observations, such a monotonic evolution is not in line, given that the recurrence time of bursts in epoch B can be even shorter than that in Epoch A (Figure~\ref{fig:recur_start}).

Our sample covers a wide parameter space of burst properties, allowing us to study the intrinsic connections between burst parameters.
Our results are summarized as follows:
\begin{itemize}
        \item We confirm the previously reported relaxation oscillator behavior $E \propto \Delta t^\alpha$ \citep[e.g.,][]{1993Lewin}, and find that the index $\alpha$ varies between mode-1 epochs (E-F) and mode-2 epochs (G-I). 
        \item To our knowledge, this is the first time that a BPL model has been used to describe the  $E-\Delta t$ relation, implying that short and long bursts correspond to different behaviours.
        \item We support the previous conclusion that the $t_{\rm dur}$–$\Delta t$ relation is suitable for describing the relaxation oscillator behavior \citep{2015.T.Bagnoli}. 
        In addition, we find that for mode-1 bursts it can also be described by a BPL model, similar to the $E-\Delta t$ relation (Figure~\ref{fig:duration-recur}).
        \item There is a clear correlation between $F_{\rm peak}$ and $E$ (Figure~\ref{fig:peak-fluence}). In addition, for 2017 bursts, it seems that there is a saturation trend, where the rise in $F_{\rm peak}$ slows significantly with increasing $E$ when $E$ is relatively large.
        \item No correlation is found between $F_{\rm peak}$ and $\Delta t$ (Figure~\ref{fig:peak-recurrence}).
        \item A linear correlation between $t_{\rm dur}$ and $E$ is found (Figure~\ref{fig:duration-Eb}).
\end{itemize}

\subsection{Comparison with previous studies}
\citet{2015.T.Bagnoli} analyzed a large sample of bursts observed with {\it RXTE}, and studied burst parameter distributions and their mutual correlations.
Our results are generally consistent with their findings, including the positive $E-\Delta t$, $t_{\rm dur}-\Delta t$, $t_{\rm dur}-E$ correlations, as well as the absence of a significant correlation between $F_{\rm peak}$ and $\Delta t$.

The main improvement of this work is that we investigated these relations in individual epochs at different outburst phases, rather than combining all bursts as a whole sample.
This allows us to study subtle features without being affected by systematic scatterings between different epochs. 
For instance, for mode-1 bursts in Epochs E and F, the $E-\Delta t$ and $t_{\rm dur}-\Delta t$ relations can be well fitted by a BPL model, significantly better than a single power-law model at significance levels of $3\sigma$ and $5\sigma$, respectively.
Actually, such a break can be also hinted if one carefully examines the large sample result (the middle panel of Figure~8 in \citet{2015.T.Bagnoli}).
As shown in Figure~\ref{fig:relaxation}, most bursts with $\Delta t> t_{\rm b}$ are classified as long bursts.
This suggests that there is some difference between long and short bursts.
In literature, the discrepancy between long and short mode-1 bursts in the $E-\Delta t$ diagram was reported by \citet{1999.R.Guerriero}.
But, unlike our results, in their samples the $E-\Delta t$ relation was observed only for long bursts and $E$ appeared to be independent of $\Delta t$ for short bursts.

On the other hand, when comparing bursts obtained from different Epochs, offsets of $E-\Delta t$ relations were found, as shown in Figure~\ref{fig:relaxation}.
Similar offsets have been reported by \citet{1999.R.Guerriero}, interpreted as the factor that the energy released (or mass accreted) per unit time decays as the outburst fades, which in turn leads to the increasing of the recurrence time to accumulate enough matter for bursts.
For 2017 bursts, we did not find such an offset, which can be explained by there are no significant changes in the accretion rate between epochs A and B.
This is plausible given the short gap (less than 2 days) between the two epochs.
In addition, in our sample the index $\alpha$ of mode-2 bursts in different epochs is always approximately 1, which is distinct from previous report by \citet{1999.R.Guerriero}, who found that the $\alpha$ value varied in the range 0.43–0.94.
The underlying reason for the $\alpha$ variation remains unclear.

For 2017 bursts, we find that $F_{\rm peak}$ increases steeply with $E$ until it reaches a critical value corresponding to a luminosity of  46\% Eddington luminosity.
\citet{2015.T.Bagnoli} also reported a saturation behavior of $F_{\rm peak}$, but this effect occurred near the Eddington luminosity.
This discrepancy may  partially originate from the systematic uncertainty introduced by the assumed broadband spectral models.
However, as mentioned above, assuming different models would only introduce a systematic error of 15\%, which is not enough to account for the discrepancy.
Therefore, we speculate that this discrepancy is more likely to reflect the intrinsic scattering of the saturation flux obtained from different observations.
For example, the saturation value discovered by \citet{2015.T.Bagnoli} was primarily dominated by mode-1 bursts, whereas the bursts we studied in 2017 were of mode-2, which usually exhibit lower $F_{\rm peak}$ values.

\subsection{Formation mechanism of the $E-\Delta t$ relation}
\label{sec:relation}
To date, the trapped-disc model is the most widely accepted physical model for type-II bursts from the RB. 
In this scenario, the neutron star's magnetic field truncates the accretion disc at the magnetospheric radius $r_{\rm m}$, which lies outside the co-rotational radius $r_{\rm co}$.
As accreted matter piles up in the inner disc region, the enhanced gas pressure drives the $r_{\rm m}$ inwards.
Once the inner disc radius crosses the $r_{\rm co}$, a rapid accretion episode begins.
During this episode, a large amount of mass falls onto the NS surface suddenly, releasing gravitational potential energy and producing a burst \citep{2010Angelo}.
This theoretical framework is supported by observations in many aspects.
For example, \citet{2015Bagnoli.b} suggested that type-II bursts are likely linked to accretion disc instabilities.
\citet{1999.R.Guerriero} reported that type-II bursts only appear at luminosities below 0.1 $L_{\rm Edd}$, corresponding to low mass accretion rates during the trapped-disc phase.
\citet{2017van} measured the inner disc radius of the RB to be $\sim$40 gravitational radii, consistent with the truncated disc scenario.
Most critically, the burst recurrence timescale predicted by this model matches the viscous timescale in the inner accretion disc, which provides the most promising agreement with the observed recurrence times of type-II bursts from the RB \citep{2015.T.Bagnoli}.
In addition, this model was also used to explain optical quasi-periodic bursts in accreting white dwarf MV Lyr \citep{Scaringi2017}, whose profiles resemble type-II bursts in the R.

This model provides a natural explanation for the observed $E$–$\Delta t$ relation. Because the greater the energy is released in a burst (i.e., the more mass accreted), the longer time is required for the inner disc to accumulate sufficient mass to overcome the magnetospheric barrier and trigger the next burst \citep{2010Angelo}.
For mode-1 bursts in Epochs E and F, we find that a BPL model is required to describe the $E$-$\Delta t$ relation, and bursts with higher fluence (long bursts) have a lower powerlaw index compared to bursts with lower fluence (short bursts).
In other words, to trigger a subsequent burst at a given fluence, the system requires a longer accumulation period after long bursts as compared to short bursts.
This may suggest that long bursts would result in a greater depletion of the trapped-disc so that a longer time is required to rebuild the pile-up region.
In addition, considering long bursts have district profiles from short bursts, their occurrence might suggest changes of the inner disc, such as variations in the thickness or temperature. 
If this is the case, a change in the $E$–$\Delta t$ relation would be expected.

This work is supported by the National Natural Science Foundation of China under grants No. 12173103 and 12261141691. 


\appendix
\section{LIGHTCURVE RECONSTRUCTION}
\label{app1}

To disentangle the contributions of two sources located in the same FoV, we adopted the method reported by \citet{Xiao2024}.
For Insight-{\it HXMT}/LE, there are three different collimators with different FoVs and orientations\footnote{The detector IDs corresponding to each collimator can be found in the documentation of the tool \texttt{hprintd\_detid}.}.
For each orientation ($i$), the observed count rate $F_{\rm i}$ can be expressed as
\begin{align}
    F_{\rm i} = F_A \times S_{\rm i,A} + F_B \times S_{\rm i,B} + B_{\rm i}
    , \quad i=0,1,2
\end{align}
, where $F_{\rm A}$ and $F_{\rm B}$ are the fluxes for the RB and the contamination source, and $S_{\rm i,A}$ and $S_{\rm i,B}$ are their corresponding effective areas, which can be calculated using the tool \texttt{lerspgen}.
$B_{\rm i}$ is the instrumental background that can be estimated by using the tool \texttt{lebkgmap}.
In practice, we extracted lightcurves $F_{\rm i} (t)$ for all three collimators with a binsize of 16\,s, and for each time bin $F_{\rm A}$ and $F_{\rm B}$ can be solved, since there are three equations with only two unknown variables.





\end{document}